\documentclass[twocolumn]{aastex62}
\usepackage{amsmath,amstext}
\usepackage{apjfonts} 
\usepackage{verbatim}
\usepackage{chngcntr}
\newcommand{\be}{\begin{eqnarray}}
\newcommand{\ee}{\end{eqnarray}}

\begin{document}

\normalsize


\title{\Large \textbf{Impact of Tides on the Potential for Exoplanets to Host Exomoons}}

\author{Armen Tokadjian}
\affiliation{Department of Physics and Astronomy, University of Southern California, Los Angeles, CA 90089-1342, USA; tokadjia@usc.edu}
\affiliation{The Observatories of the Carnegie Institution for Science, 813 Santa Barbara St., Pasadena, CA 91101, USA}

\author{Anthony L. Piro}
\affiliation{The Observatories of the Carnegie Institution for Science, 813 Santa Barbara St., Pasadena, CA 91101, USA}

\begin{abstract}
Exomoons may play an important role in determining the habitability of worlds outside of our solar system. They can stabilize conditions, alter the climate by breaking tidal locking with the parent star, drive tidal heating, and perhaps even host life themselves. However, the ability of an exoplanet to sustain an exomoon depends on complex tidal interactions. Motivated by this, we make use of simplified tidal lag models to follow the evolution of the separations and orbital and rotational periods in planet, star, and moon systems. We apply these models to known exoplanet systems to assess the potential for these exoplanets to host exomoons. We find that there are at least 36 systems in which an exoplanet in the habitable zone may host an exomoon for longer than one gigayear. This includes Kepler-1625b, an exoplanet with an exomoon candidate, which we determine would be able to retain a Neptune-sized moon for longer than a Hubble time. These results may help provide potential targets for future observation. In many cases, there remains considerable uncertainty in the composition of specific exoplanets. We show the detection (or not) of an exomoon would provide an important constraint on the planet structure due to differences in their tidal response. 
\end{abstract}

\keywords{exoplanets: exomoon ---
		exoplanets: tides ---
		habitability }

\section{Introduction}
Over 4000 exoplanet candidates have been discovered in the past few decades, yet there has been no confirmed exomoons (although see \citealp{teachey} for a possible detection). Using the Earth's moon as a standard, in our own solar system there are five moons at least as massive, and several within an order of magnitude. The abundance of moons in our neighborhood strongly suggests that exomoons will eventually be found to be ubiquitous in most exoplanet systems.

The presence of a moon is not just a curiosity, since it may play an important role in the habitability of the hosting planet. A moon with mass similar to Earth's moon or larger may give rise to a number of significant effects. It can stabilize the obliquity of a planet so that the axial tilt does not vary erratically \citep{laskar1993}. This prevents extreme temperature swings at the surface of a planet, which might be crucial for some organisms, depending on their adaptability. Additionally, a moon may break the tidal locking of a planet to its star, thereby preventing a day-only and night-only hemispheric split on the planet. As shown in \citet{Piro18}, a moon also drives tidal heating within a planet. This helps power tectonic activity in the planet crust, which aids in the movement of gases in the carbon cycle and may play a role in the prevention of the runaway greenhouse effect surmised to have occurred on Venus \citep{jackson}. A large enough moon may also prevent the cooling of the planet core, thereby sustaining the dynamo effect that drives the magnetic field \citep{andrault}. The presence of a magnetic field blocks harmful radiation and preserves the atmosphere, in contrast to the fate of Mars. Finally, for those planets that do not have a solid surface, life has the potential to develop on the moon itself.

Given the importance that moons may have, we explore the potential for exoplanets to host exomoons. Using parameterized models for the tidal dissipation in the planet, we follow the spins and orbital separation evolution for planet-moon-star systems. These are applied to known Kepler exoplanets to assess the length of time they could potentially host a moon. This is an extension of previous work by \citet{Martinez} which looked into the dynamical stability of exomoons around M dwarf exoplanets, but here we compute the full time evolution to understand the moon survival times of single-planet systems. We highlight exoplanets within the habitable zone of their host star which could hold a moon for over approximately a gigayear, the rough timescale for the development of life \citep{Schopf2017}. Many current exoplanet candidates lack a mass measurement, but we demonstrate that the ability for their exoplanets to hold a moon for a significant time depends strongly on the strength of their tidal dissipation. Thus, if these planets were found to host moons, it would provide important information about the planet's structure.
 
In Section~\ref{sec:equations}, we summarize the planet-moon separations allowed in these systems, and introduce the two tidal lag models used in our analysis. In Section~\ref{sec:theo apply}, we explore the general properties of these models to show the dependence on different parameters. In Section~\ref{sec:real apply}, we apply the models to real exoplanets, and then discuss notable specific systems in Section~\ref{sec:results}. We describe the consequences and future implications of this study in Section~\ref{sec:future} and provide a summary in Section~\ref{sec:conclusion}.

\section{Orbital Dynamics}
\label{sec:equations}

We first summarize the basic framework we use for studying the tidal interactions for a system of a single planet, a moon, and a star. Such systems have been addressed in a number of previous studies (e.g., \citealp{ward}; \citealp{touma}, \citealp{neron}; \citealp{barnes2002}; \citealp{sasaki12}; \citealp{sasaki14}; \citealp{adams2016}). Here we follow the basic picture presented in \citet{Piro18}. Considering just the planet and the moon, the moon generates two bulges on the planet, one on the close side and one on the far side. However, the planet cannot react instantaneously to the tidal forcing. Thus, the bulge will not directly align with the position of the moon, but will be offset, depending on the composition, rigidity, and dissipation rate of the planet. This is referred to as a ``tidal lag''. If the planet and moon are not tidally locked, the planet will spin faster (or slower) than the moon is orbiting. Thus, the moon will essentially be lagging (or leading) the bulge, causing a net torque that will slow down (or speed up) the rotation of the planet. This in turn will cause the moon to spiral outwards (or inwards).

In the following subsections, we introduce the governing equations that describe this process. We then show the limits on the planet-moon separation parameter and detail the two tidal lag models used in our analysis.

\subsection{Governing Equations}
To solve for the time evolution of the planet-star-moon systems, we use a set of coupled differential equations describing the rotational frequency $\Omega_p$ of the planet, the planet-moon separation $a_m$ and the planet-star separation $a_s$. The resulting three differential equations describe the spin evolution of the planet due to the combined torques from the star $N_s$ and the moon $N_m$,
\be
	\frac{d}{dt}(I_p\Omega_p)=N_s + N_m,
	\label{spin}
\ee
the orbital evolution of the planet, 
\be
	\frac{d}{dt}[M_p(GM_sa_s)^{1/2}]=-N_s,
	\label{seps}
\ee
and the orbital evolution of the moon,
\be
	\frac{d}{dt}[M_m(GM_pa_m)^{1/2}+I_m n_m]=-N_m.
	\label{sepm}
\ee
The masses of the planet, moon, and star are $M_p$, $M_m$, and $M_s$, respectively and the moment of inertia of planet and moon are $I_p$ and $I_m$ respectively. The moment of inertia of the star is not included because we assume that the inertia is relatively large so that the star's spin is insignificantly altered by the tides of the planet. While this approximation may break down for hot Jupiters whose masses are much closer to the star mass and are in short period orbits, it does not affect the habitable zone planets which is the main focus of this paper. The orbital frequency of the moon is denoted as $n_m$ and the term $I_mn_m$ in Equation~(\ref{sepm}) assumes that the moon is tidally locked to the planet because of the relatively short locking time of the moon, which is on the order of millions to tens of millions of years for the Earth-Moon system. Whether or not the moon is actually tidally locked to the planet has little effect on our main results given the small spin angular momentum of the moon.

This set of equations makes a number of simplifying assumptions. All orbits are taken to be circular, thus we do not include the rate of change of eccentricity. In addition, we consider secular timescales so that the tidal bulges caused by the moon are torqued by only the moon and tidal bulges caused by the star are torqued only by the star and not vice versa. This is because over many orbits the torque from one body on the bulge caused by another will cancel. Although giant planets may undergo significant structural evolution which may impact a moon's tidal evolution at early times \citep{alvarado}, for this work we assume the moon survives these initial phases. Finally, we assume that the spin and orbital angular momenta are all aligned. 

\subsection{Moon Separation Limits}
Here we summarize how we set the initial semi-major axis for the moon. The outer limit depends on the distance between planet and star, as the closer in the planet is to its star, the more likely the star is to strip the moon away from the planet. The maximum separation between planet and moon also depends on the ratio between planet mass and star mass, and is given by

\be
	a_{m,\mathrm{max}} = fa_s\left(\frac{M_p}{3M_s}\right)^{1/3}.
	\label{ammax}
\ee A constant value of 0.49 is used for $f$, as detailed in \citet{Domingos2006}. For a system in which the moon's orbital rate is less than the planet's spin rate, the moon will slow down the planet's rotation and move further out in its orbit (as is the case currently for the Earth). Should the moon exceed this maximum distance, it will be lost to the parent star.

On the other hand, there is also a minimum separation required for stability between planet and moon. This is set by where the planet's tidal forces on the moon exceed the moon's self gravity, the so-called Roche lobe. This inner boundary is thus dependent on the radius of the moon and the ratio between planet mass and moon mass \citep{Frank2002}

\be
	a_{m,\mathrm{min}} = 2.16 R_m\left(\frac{M_p}{M_m}\right)^{1/3}.
	\label{ammin}
\ee 
In detail, the moon may be subject to tidal disruption or send it colliding with the planet as it reaches this semi-major axis. Indeed, the ultimate fate depends on the relative densities between planet and moon as discussed in more detail in \citet{Piro18}. For this work, we are mainly focused on the lifetime of the moon itself and not on the details of its final fate. 

\subsection{Tidal Lag Models}
As celestial bodies exert tidal forces on one another, their surfaces and interiors are squeezed and pulled, causing deformations that are eventually dissipated. This tidal dissipation is a complex problem explained by a variety of proposals. For instance, \citet{Ogilvie} details the process as a fluid dynamics problem caused by a coupling of the gravitational quadrupole moment of one body with the monopole of the other. Other models include dissipative creep processes in rocky material \citep{storch} and excitation of oscillatory modes in gas giants \citep{vick}.

Due to the complicated process involved in the dissipation of tides, we consider parameterized models to encapsulate the main impact of tidal interactions. We consider two different formulations for this to better show the range of uncertainty inherent in this approach. Although we find quantitative differences between the different parameterizations, our results are qualitatively the same in each case.

The first is the constant time lag (CTL) model. In this model, we take the delay between gravitational pull and bulge appearance to be a constant time, depending on the structure of the planet. In this case, the torque is proportional to the time lag $\tau$ of the planet and the difference in the rotational frequency of the planet and orbital frequency of the planet around the star or moon around the planet. Indeed, since the time it takes for the bulge to appear is constant, the faster the planet rotates relative to how fast the moon orbits the larger the torque will be. Conversely, as the planet and moon (or planet and star) approach tidal locking the torque shrinks and approaches zero. The equations for the torques detailing this process are \citep{Ogilvie}

\be
	N_s = 3k_2\tau(n_s-\Omega_p)\frac{GM_s^2R_p^2}{a_s^6},
	\label{torques}
\ee
\be
	N_m = 3k_2\tau(n_m-\Omega_p)\frac{GM_m^2R_p^2}{a_m^6}.
	\label{torquem}
\ee
Here $k_2$ is the Love number of the planet that depends on its rigidity. We take this $k_2$ to be 0.3 for rocky planets like the Earth \citep{Yoder1995}, 0.38 for gas giants like Jupiter \citep{Gavrilov}, and 0.34 for ice giants like Neptune (see the interval described in \citealp{Kramm}). Furthermore, the time lag $\tau$ is taken to be 638\,s \citep{lambeck,neron} for rocky planets, 0.766\,s for Neptune-like planets, and 0.00766\,s for Jupiter-like planets. The basis for timescales in the latter two is from taking ratios between the quality factor $Q$ described in the constant phase lag (CPL) model (see below).

\begin{figure}
\epsscale{1.2}
\plotone{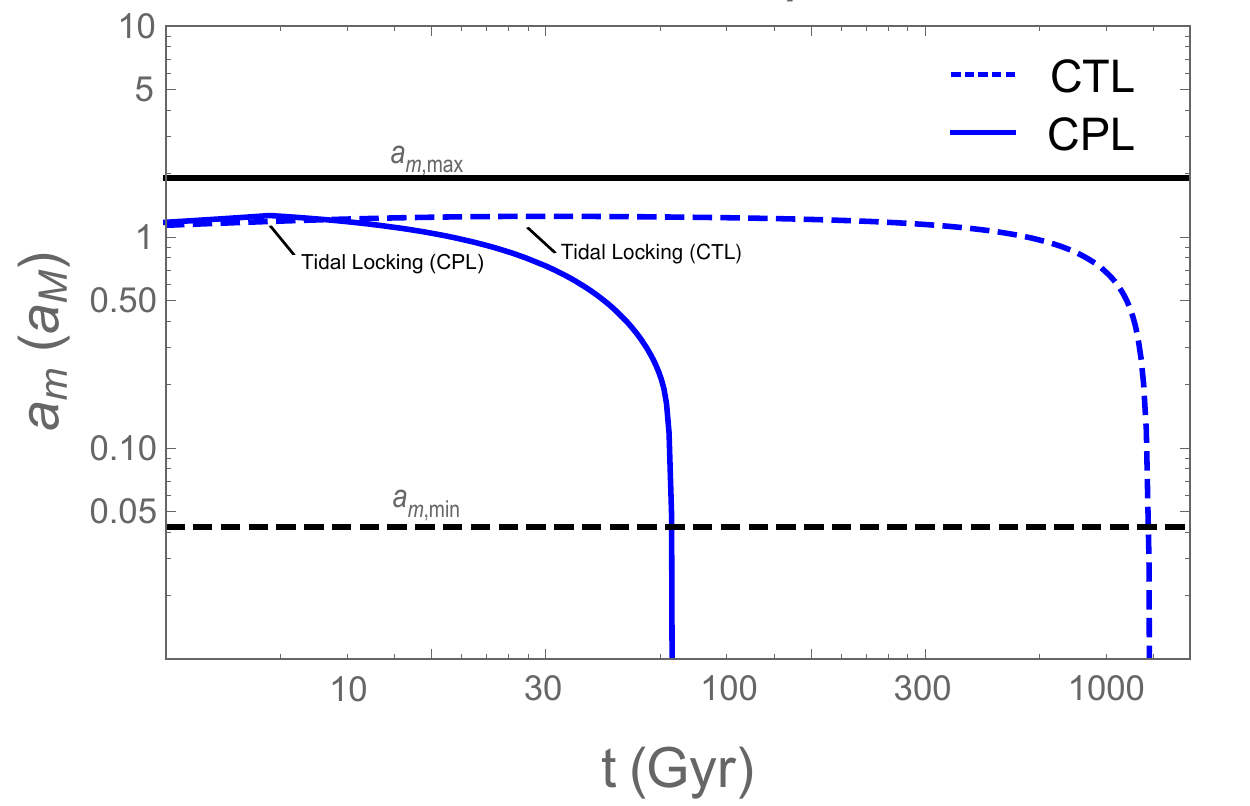}
\caption{The planet-moon separation in terms of $a_M$, the current Earth-Moon separation, of an Earth-like system over time for the CTL and CPL model. The solid horizontal line represents $a_{m,\mathrm{max}}$, while the dashed horizontal line is $a_{m,\mathrm{min}}$. In the CTL model, the Moon is tidally disrupted after over 1000\,Gyr whereas in the CPL model, it is disrupted after only 68\,Gyr.}
\label{CTLearth}
\epsscale{1.0}
\end{figure}

An example of applying the CTL model is shown in Figure~\ref{CTLearth}. Here we start with the present Earth, Sun, and Moon system and with these initial conditions, we integrate the coupled differential equations (\ref{spin}), (\ref{seps}), and (\ref{sepm}) forward in time and plot the resulting Earth-Moon separation. For simplicity for illustrative purposes here, we have excluded all other solar system bodies. In this system, Earth is initially rotating faster than the Moon is orbiting. This causes the Moon to torque the Earth down and the Moon starts out moving away until eventually it becomes tidally locked to the Earth. At this point, the torque due to the Sun becomes dominant, slowing the Earth down further. Thus the Moon torques up the Earth and spirals inwards. The process continues and leads to a runaway effect of the Earth spinning faster and faster and the Moon moving in closer and closer until it is tidally disturbed after about 1250\;Gyr. At this point, the Earth would have spun up to have a a day equal to about 6 hours, which is the thought to be early Earth's rotational period.

The second tidal lag model we use is the CPL model. In this model, the planet responds to the gravitational perturbation at a constant phase rather than after a constant time. This means that the angle between the resultant bulge and the line joining the centers of the planet and moon (or star) will always be the same. In contrast to the CTL model, the torque in this model is constant, dependent only on the quality factor $Q$ and the sign of the difference between rotational and orbital frequency. 

The quality factor is a measure of the efficiency of tidal dissipation of the planet and is inversely proportional to the angular lag \citep{Efroimsky}. So, the higher the $Q$ the smaller the angle and less dissipative the planet. In fact, rocky planets like the Earth have a $Q$ of about 12, whereas ice giants like Neptune are much higher at $10^4$ and Jupiters higher still at $10^6$ \citep{murray,goldreich}. 

The time lag $\tau$ and quality factor $Q$ are related simply by $2 k_2 \tau (n_s - \Omega_p) = k_2 \sigma_s /Q $, where $\sigma_s = \rm {sgn}(n_s-\Omega_p)$ (and similarly $n_m$ and $\sigma_m$ for the moon). Thus in the CPL model the torque magnitude is dependent only on $Q$ and the direction is given by the rotational and orbital frequencies. The torques are given by \citep{efroimsky13}

\be
	N_s = \frac{3}{2}\frac{k_2}{Q}\sigma_s\frac{GM_s^2R_p^2}{a_s^6},
	\label{CPLtorques}
\ee
and
\be
	N_m = \frac{3}{2}\frac{k_2}{Q}\sigma_m\frac{GM_m^2R_p^2}{a_m^6}.
	\label{CPLtorquem}
\ee

To summarize, the torque is constant in the CPL model and dependent on the difference between rotational frequency of the planet and orbital frequency of the moon around the planet in the CTL model. As the moon slows down or speeds up the planet rotation towards synchronous rotation with the moon's orbit, the torque get weaker and weaker in the CTL model. For this reason, the CTL model tends to produce longer timescales for moon survival. We show the same Earth-Moon-Sun model in Figure~\ref{CTLearth}, this time using the CPL model. As shown, the Moon now lasts a much shorter 68\,Gyr before becoming tidally disrupted.

\section{Theoretical Systems}
\label{sec:theo apply}

\begin{figure}
\epsscale{1.2}
\plotone{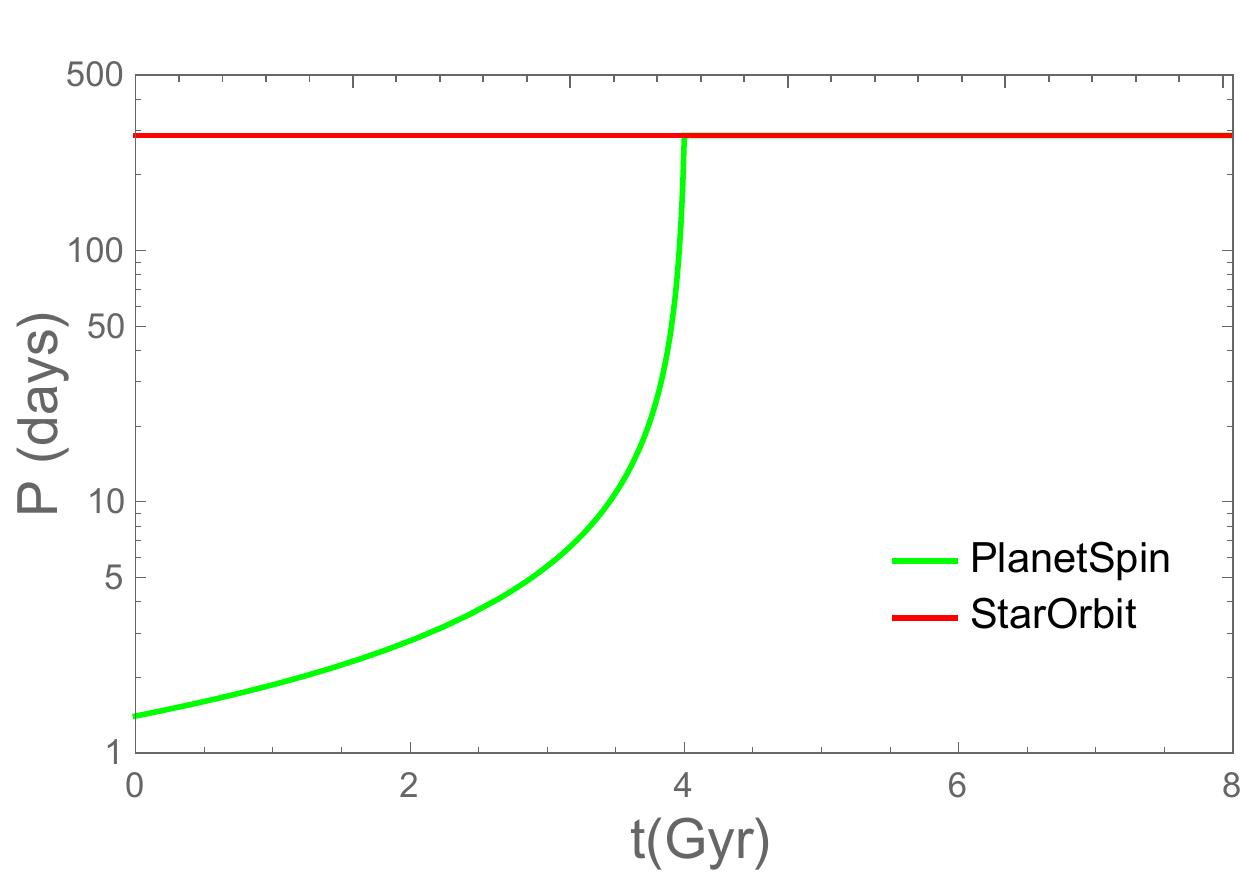}
\caption{A single planet system with no moon that will become tidally locked to its parent star in about 4\,Gyr. The green curve is the rotational period of the planet and the red curve is the orbital period of the planet around the star.}
\label{lock}
\epsscale{1.0}
\end{figure}

We next apply our model to theoretical planet, star, and moon systems to explore the general trends we expect them to exhibit. We first show that a planet tidally locked to its star can be ``unlocked'' by the presence of a moon. Consider a single rocky planet with no moon that is tidally locked to its star, that is, its rotational frequency is equal to its orbital frequency around the parent star. We will use the CPL model for this example, although the CTL model gives qualitatively similar results. Figure~\ref{lock} shows such a two body system: a planet of mass $M=1.3M_{\oplus}$ and radius $R=1R_{\oplus}$ with initial rotational period of 1.4 days orbits a star of mass $M_s=1M_{\odot}$ at an initial distance of 0.85\,AU and becomes locked to it by approximately 4\,Gyr. The plot shows the periods (in days) rather than frequency to be more intuitive. If such a system is left alone, the planet stays tidally locked to its star. 

\begin{figure}
\epsscale{1.2}
\plotone{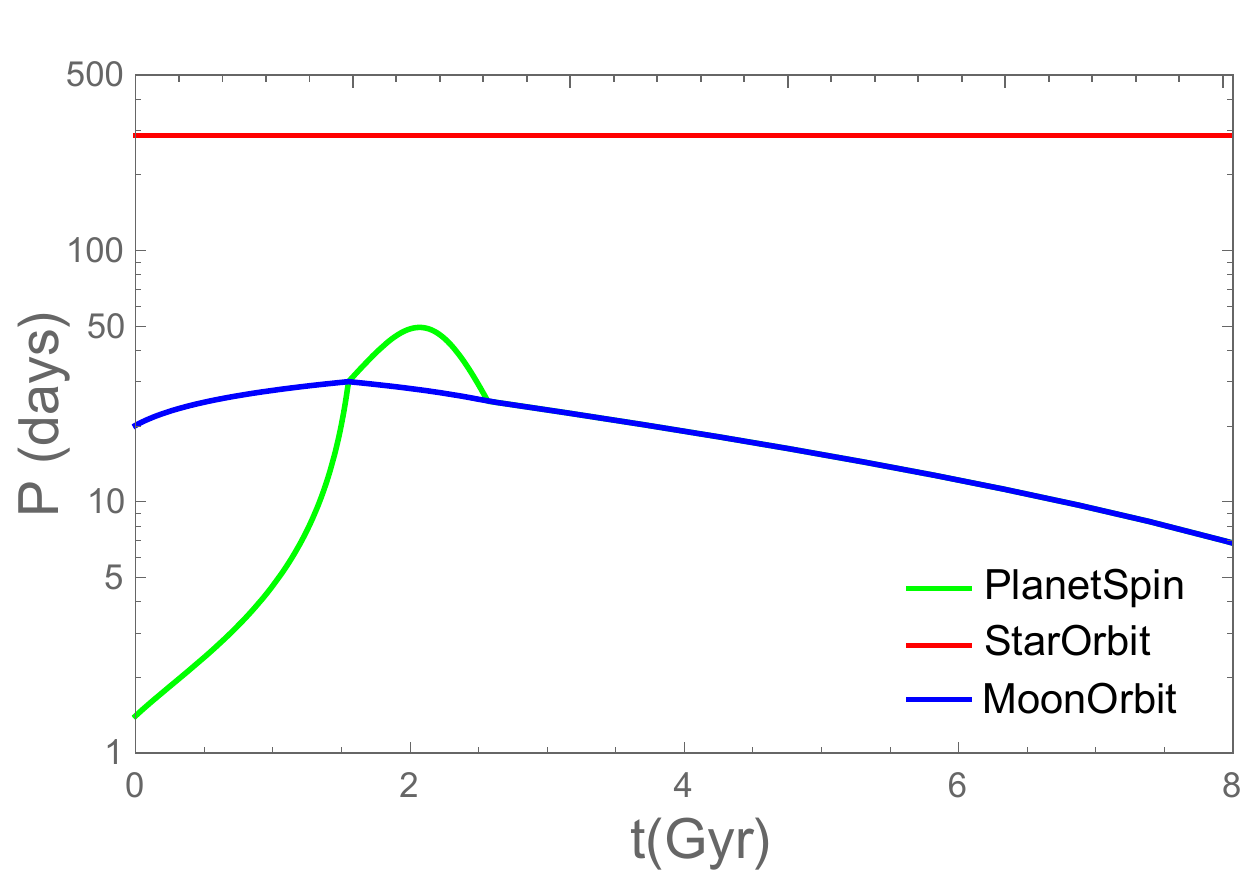}
\caption{The same system as in Figure~\ref{lock}, this time with the presence of a moon. In this case, the planet does not lock to its star. The blue curve is the orbital period of the moon around the planet, which oscillates about the rotational period of the planet in a quasi-tidally locked configuration.}
\label{breaklock}
\epsscale{1.0}
\end{figure}

Now we add a moon with the mass and radius of our moon and place it at the midpoint of the minimum and maximum allowed planet-moon separations. We call this separation $a_{{m,\rm{avg}}}$. Figure~\ref{breaklock} shows the evolution of the now 3-body system through time. Due to the tidal effects of the moon, the planet does not lock to its star in this case, but rather temporarily becomes locked to the moon after about 3\,Gyr. The planet and moon are not actually locked at this time but follow a sequence of the moon coming in closer and the planet spinning faster until eventually the moon comes close enough to the planet to be disrupted.  

The climate on a planet is likely very different between a case where it is tidally locked to the star, which would result in one side always being hot and the back side always being cold, and a case where the planet has day/night cycles. This in turn may impact which planets are expected to be favorable for hosting life. In fact, tidally locked planets are more likely to initiate a runaway greenhouse effect or incur atmospheric collapse \citep{kite}, and they are less likely to go through a `snowball planet' phase which is thought to be an important step for the onset of life \citep{checlair}.

There are a plethora of variables that play a role in the tidal evolution models we employ. The parameters of the moon are perhaps the least important. The inner limit of planet-moon separation depends on moon density, which does not vary much when noting the moon densities in our solar system. So for this work we take all moon densities to be $3\,\rm{g}\,\rm{cm}^{-3}$. The torque due to the moon on the planet's bulge depends on moon mass, so to ensure the moon has an influence we take a modest sized moon and set the mass of all moons henceforth to be the same as our Moon's, $1\,M_M$.

Planet-moon and planet-star separations are crucial parameters, as indicated by the sixth power in Equations (\ref{torques}) and (\ref{torquem}). The planet-star separation for exoplanets is relatively well constrained, and the planet-moon separation is a free variable we can choose since we insert an artificial moon in our calculations. Thus, it is important to consider other parameters when choosing an exoplanet list.

\begin{figure}
\epsscale{1.2}
\plotone{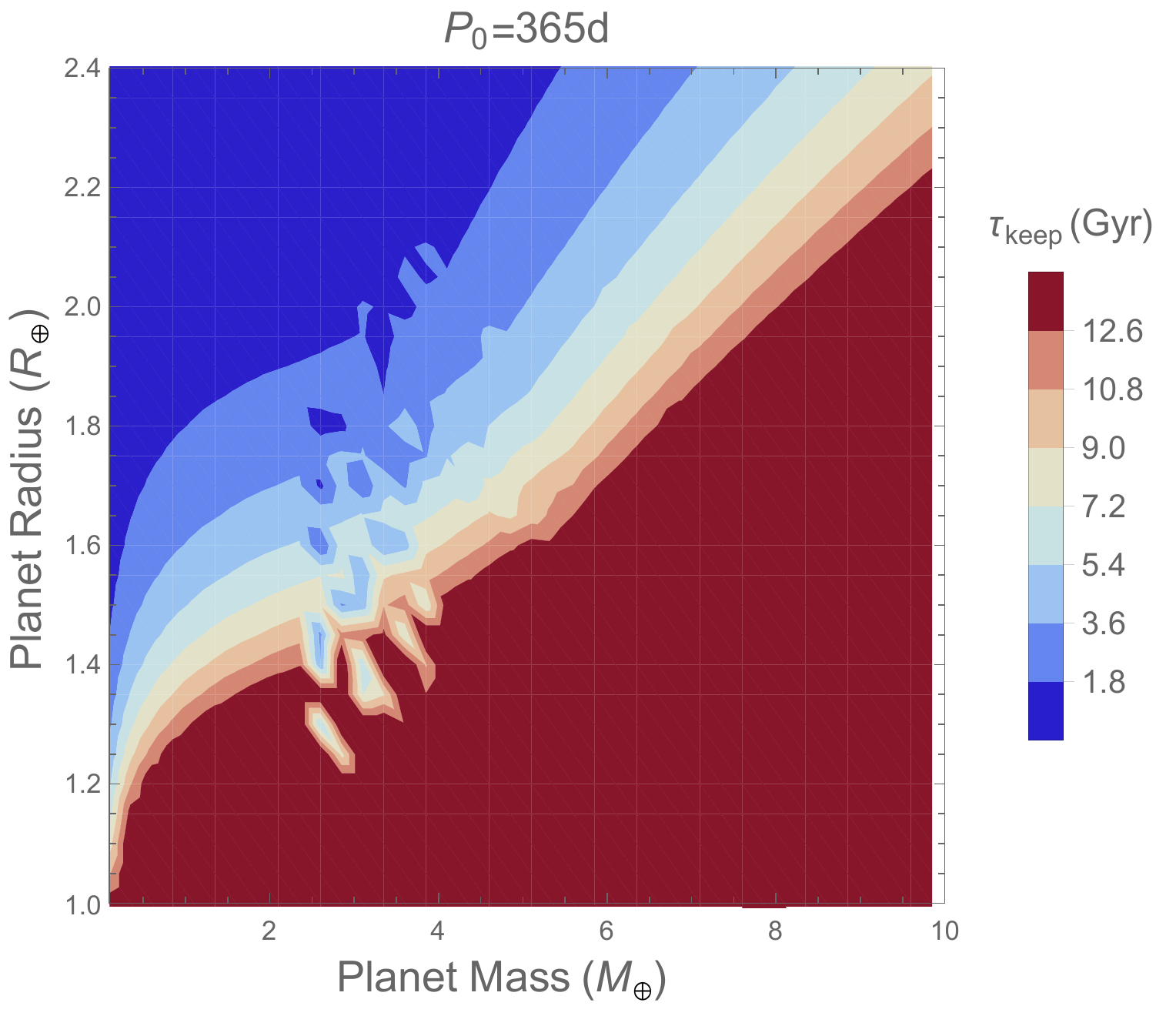}
\caption{A contour plot representing the moon keep time for a rocky planet (Q=12) of certain mass and radius around a sun-like star at 1\,AU using the CPL model. The planet is initially tidally locked to its star in this case and has a fixed density of $5\,\rm{g}\,\rm{cm}^{-3}$. The general trend is an increase in moon survivability with increasing planet mass and decreasing planet radius.}
\label{contour365}
\epsscale{1.0}
\end{figure}

\begin{figure}
\epsscale{1.2}
\plotone{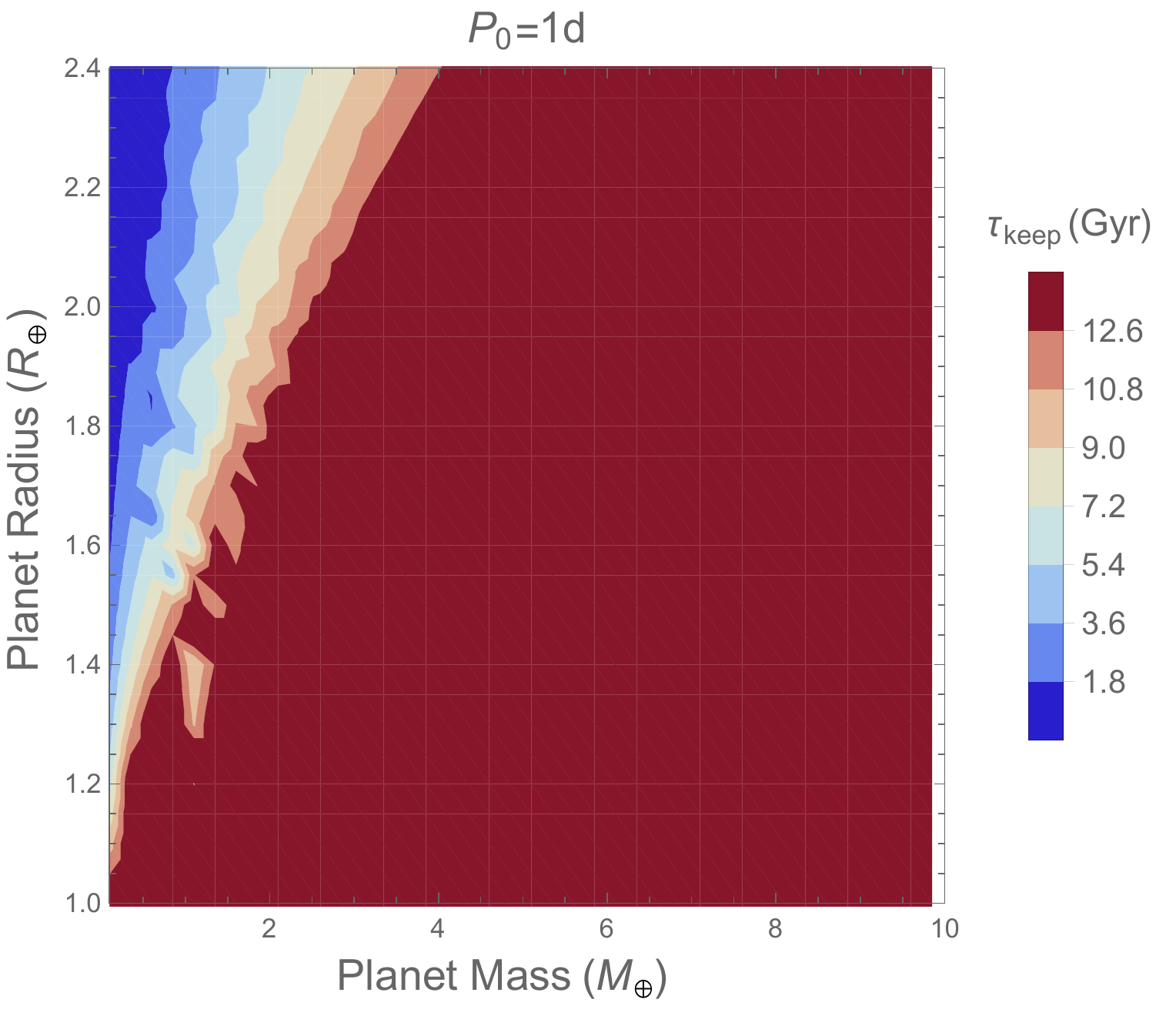}
\caption{The same as Figure~\ref{contour365}, but with an initial planet rotational period of 1 day. Overall, the timescales have increased but the same general trends apply.}
\label{contour1}
\epsscale{1.0}
\end{figure}

\begin{figure}
\epsscale{1.2}
\plotone{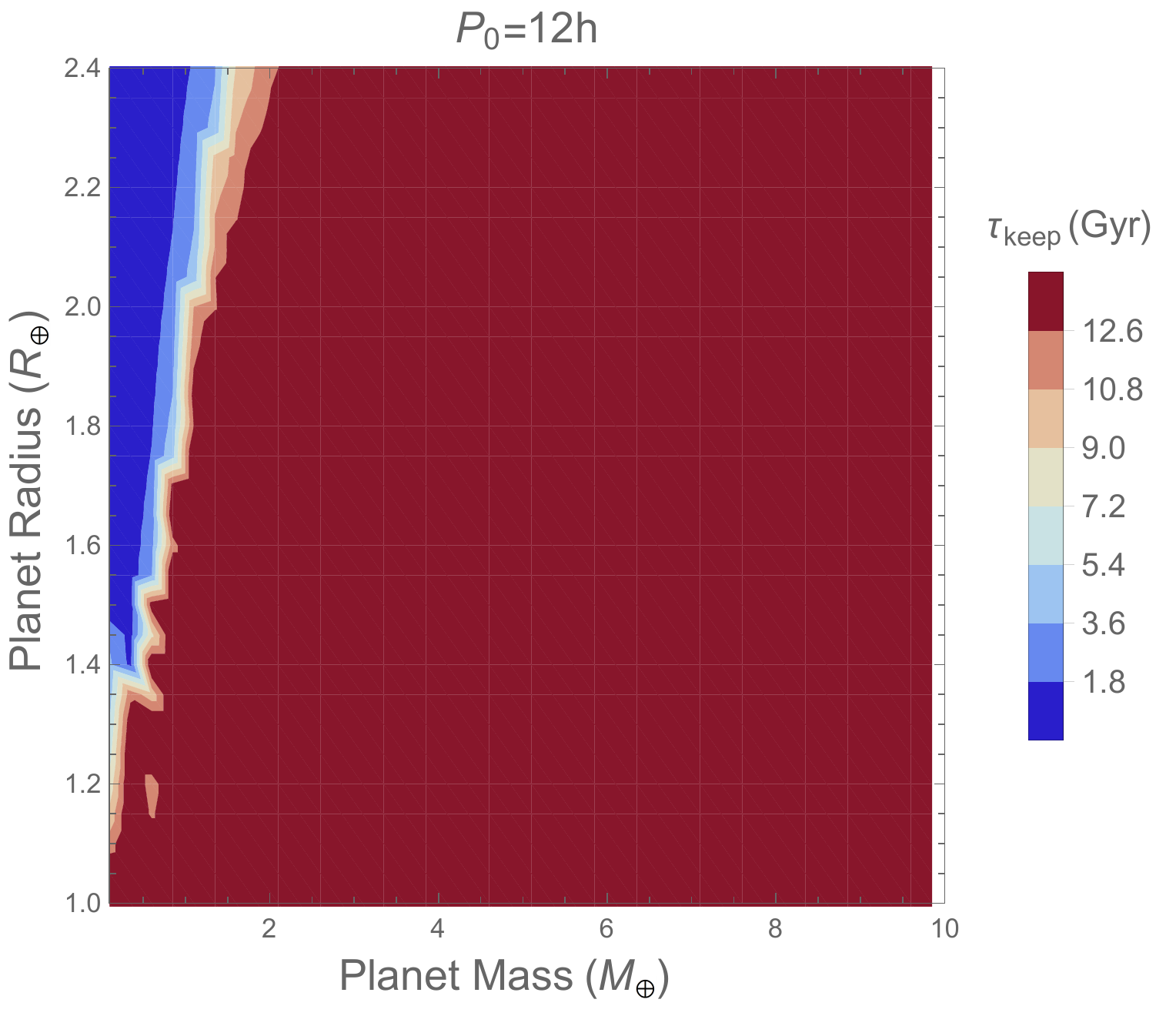}
\caption{The same as Figure~\ref{contour365}, but with an initial planet rotational period of 12 hours. Again, the timescales increase since the planet starts out with a faster rotation.}
\label{contour12h}
\epsscale{1.0}
\end{figure}

To get a sense of which variable is next most important, we make a series of contour plots comparing how long a moon is kept in a 3-body system with varying planet mass and radius. We choose the planet to be rocky with density $5\,\rm{g}\,\rm{cm}^{-3}$ and use the CPL model with $Q=12$. The mass of the star and initial planet-star separation are taken to be $1\,M_{\odot}$ and 1\,AU, respectively, for illustrative purposes. We define $\tau_{\mathrm{keep}}$ as the timescale the moon is kept by the planet before the moon is lost to the star or falls into the planet. Figures (\ref{contour365}), (\ref{contour1}), and (\ref{contour12h}) are contour plots corresponding to $\tau_{\mathrm{keep}}$ for different initial rotation period $P_0$ of the planet. The noise in each of these plots is due to undersampling of the parameter space and is not physical, but the general trends are still evident. In Figure~\ref{contour365}, $P_0$ is 365 days, so the planet is initially locked to its star. In the next two figures we decrease the periods to 1 day (Figure \ref{contour1}) and 12 hours (Figure~\ref{contour12h}). The results show that the planet radius has a strong impact on the evolution. As we increase radius vertically up on each contour at any constant mass, there is a significant decrease in the time the moon is kept in the system. In contrast, although increasing mass at constant radius increases timescales, the effect is much less significant. For example, in Figure~\ref{contour1} we can take a planet with $M=1M_{\oplus}$, $R=1.6R_{\oplus}$, which will hold onto its moon for about 8\,Gyr according to the contour plot. If we multiply the mass by 1.5 and keep the radius constant ($M=1.5M_{\oplus}$, $R=1.6R_{\oplus}$), the timescale increases to around 12\,Gyr, a 50\% increase. Now if we instead multiply the radius by 1.5 and keep the mass constant ($M=1M_{\oplus}$, $R=2.4R_{\oplus}$), then the timescale drops to under 2.5\,Gyr, a near 70\% decrease. 

The initial rotational period of the planet is important as well, but given the difficulty of constraining exoplanet rotational frequency, we take each planet to be initially locked to its star (following Figure~\ref{contour365}) for the rest of this work. Although many known exoplanets are probably not tidally locked to their star, this approximation is a conservative approach since for most cases the moon retention time is actually longer for faster initial rotation rates for the planet. This is because as long as the moon is not lost to the star, there is a longer period of time where the moon is being pushed away, and then the moon has to travel a longer distance coming back in towards the planet. Also, fast rotating Jupiter-like planets have enough angular momentum in principle to push a moon away until it is lost, but given the high mass of such planets, in practice we find that because the tidal forces are relatively weak, it leads to long moon keep times.
\section{Kepler Exoplanets}
\label{sec:real apply}

We now apply our models to real exoplanet systems filtered from the NASA Exoplanet Archive \footnote{http://exoplanetarchive.ipac.caltech.edu/}. We choose Kepler exoplanets, but since often the mass is unknown and hence the density also, the composition of many of these planets is uncertain. For this reason, we create a planet classification system based mainly on radius as shown in Table \ref{table}. We first perform a radius cut and split the planets into three groups: rocky Earth-like planets, Neptune-like planets, and Jupiter-like planets. A second density cut is utilized for radii between $1.75R_{\oplus}$ and $2.4R_{\oplus}$ to differentiate super-Earths and sub-Neptunes. The `systems' column displays the number of single planet systems that follow the criteria. 

Since for many planets the mass and density are unknown, we allow planets to be treated as two different compositions. For example, the exoplanet Kepler-1638b has a radius of $1.87\,R_{\oplus}$, falling under the radius category between $1.75\,R_{\oplus}$ and $2.4\,R_{\oplus}$, and it has unknown mass. Thus we consider the mass to be $5.9\,M_{\oplus}$ corresponding to a density of $5\,\rm{g}\,\rm{cm}^{-3}$ and treat it as a Rocky planet (super-Earth) \emph{and} we also consider the mass to be $1.8\,M_{\oplus}$ corresponding to a density of $1.5\,\rm{g}\,\rm{cm}^{-3}$ and treat it as a Neptune-like planet (sub-Neptune). (Jupiter-like planets with unknown mass have densities set to $1\,\rm{g}\,\rm{cm}^{-3}$). These densities were chosen because they are the average densities of the corresponding planet type in our solar system. Allowing for this overlap is important because composition affects moon survivability due to different $\tau$ values (for CTL) or different $Q$ values (for CPL). This is significant for two reasons. First, a planet treated as one type may not hold a moon for a relatively long enough time for life to develop whereas the same planet treated as another type may do so. Thus we increase the likelihood of finding habitable zone planets that meet the criterion of moon survivability over 1\,Gyr. Second, if an exomoon is found in the future, we can infer the composition of its host planet by looking at the age of the system and comparing it to the expected moon survivability time. With the overlap considered, we apply our models to a total of 813 Rocky planets, 773 Neptune-like planets, and 605 Jupiter-like planets.

\begin{deluxetable*}{cccccccc}
\tablecolumns{8} \tablewidth{900pt}
\tablecaption{Planet Classification Scheme and Tidal Parameters}
\tablehead{
Radius $(R_\oplus)$&Density $(\rm{g\,cm}^{-3})$&Type&Systems&Q&$k_2$&$\tau (s)$&Reference
}
\startdata
<1.75 & unknown & Rocky & 482 & 12 & 0.30 & 638 & (1)\\
1.75 - 2.4 & >3.6 & Rocky & 331* & 12 & 0.30 & 638 & (2,3,4,5)\\
1.75 - 2.4 & 2.1 - 3.6 & Rocky or Neptune & 329* & 12 or $10^4$&0.30 or 0.34&638 or 0.766&(2,3,4,5)\\
1.75 - 2.4 & <2.1 & Neptune & 331* & $10^4$ & 0.34 & 0.766 & (2,3,4,5) \\
2.4 - 3.5 & unknown & Neptune & 346 & $10^4$ & 0.34 & 0.766 & (1)\\
3.5 - 5.7 & unknown & Neptune or Jupiter & 99 & $10^4$ $\rm{or}$ $10^6$&0.34 or 0.38&0.766 or 0.00766&(6)\\
>5.7 & unknown & Jupiter & 506 & $10^6$ & 0.38 & 0.00766 & (6)
\enddata
\tablecomments{(1)~\citet{fulton2017}, (2)~\citet{valencia2007}, (3)~\citet{fortney2007}, (4)~\citet{charbonneau2009}, (5)~\citet{odrzywolek2016}, (6)~\citet{stern2002}; *from an overlap of 333 unique exoplanets}
\label{table}
\end{deluxetable*}

Our main result involves running the models on these real exoplanet systems by inserting an Earth's moon-sized moon at $a_{m,\rm{avg}} =\frac{1}{2} (a_{m,\rm{min}}+a_{m,\rm{max}})$ in each system, integrating the tidal evolution equations forward in time, and then plotting the moon survival time in Gyr against the stellar flux received by the planet in solar units (SU). We use the radius, semimajor axis, and when available, the mass of the planet when running the models. When there is no mass or only an upper limit we assign a mass based on the density desired as explained earlier. We use the stellar properties available to calculate the flux received by the planet. Finally, we assume every planet starts out tidally locked to its star, so that its initial rotation period is equal to its orbital period.

\begin{figure}
\epsscale{1.2}
\plotone{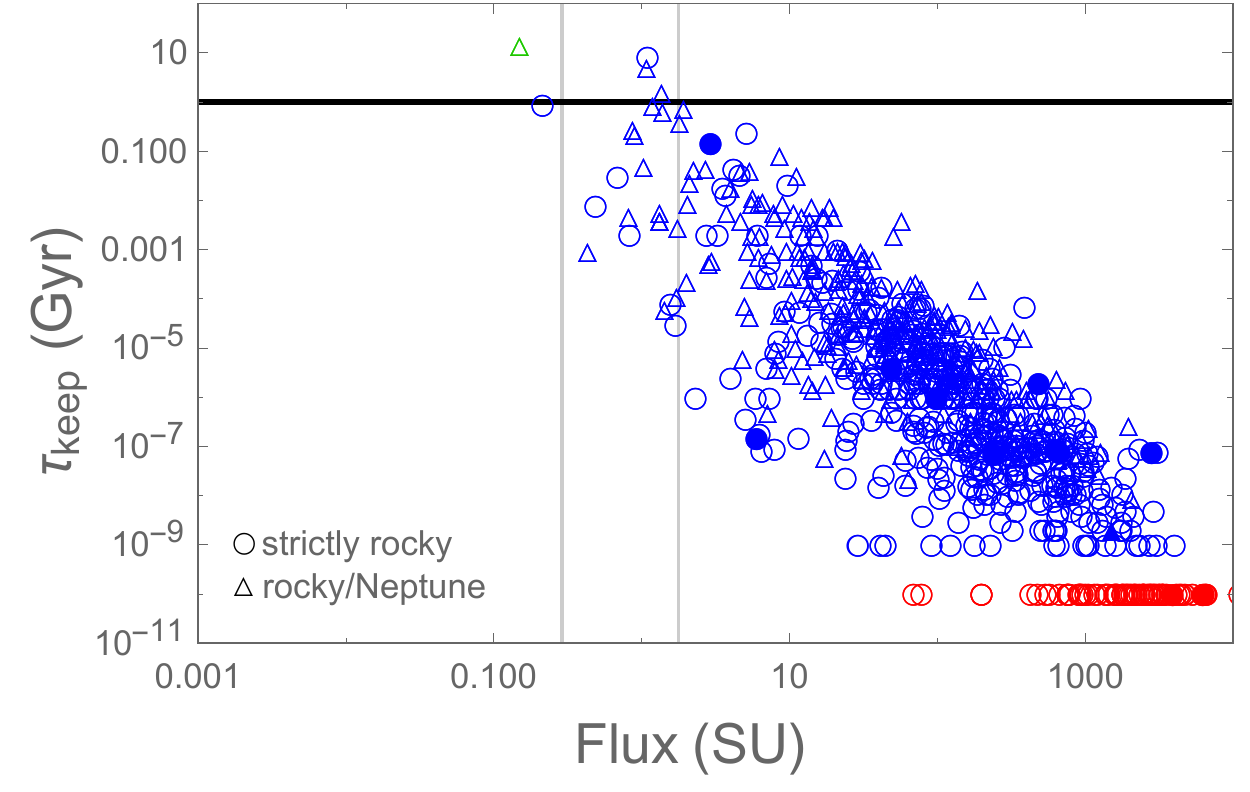}
\caption{Scatter plot showing moon keep time for real exoplanets we classify as "rocky" using the CPL model. Circles are rocky only planets and triangles are planets we consider can be both rocky or Neptune-like. Color represents moon fate: red is lost to the star, blue is tidal disruption, and green is kept over 15\,Gyr. Filled symbols are planets that have a mass estimate while open ones do not.}
\label{rockycpl}
\epsscale{1.0}
\end{figure}

Figure~\ref{rockycpl} shows the first of our results, the CPL model applied to Rocky exoplanets. The dark horizontal line is at 1\,Gyr, the approximate time required for life to develop on a planet. The two gray vertical lines mark the optimistic habitable zone for G-type stars as described in \citet{kopparapu2013}, ranging from 0.29\,SU to 1.78\,SU. Thus we are interested in the small rectangle formed by these lines in the upper middle part of the plot. The planets lying in that rectangle are those that are in the habitable zone of their respective systems and can hold a moon for at least 1\,Gyr. 

\begin{figure}
\epsscale{1.2}
\plotone{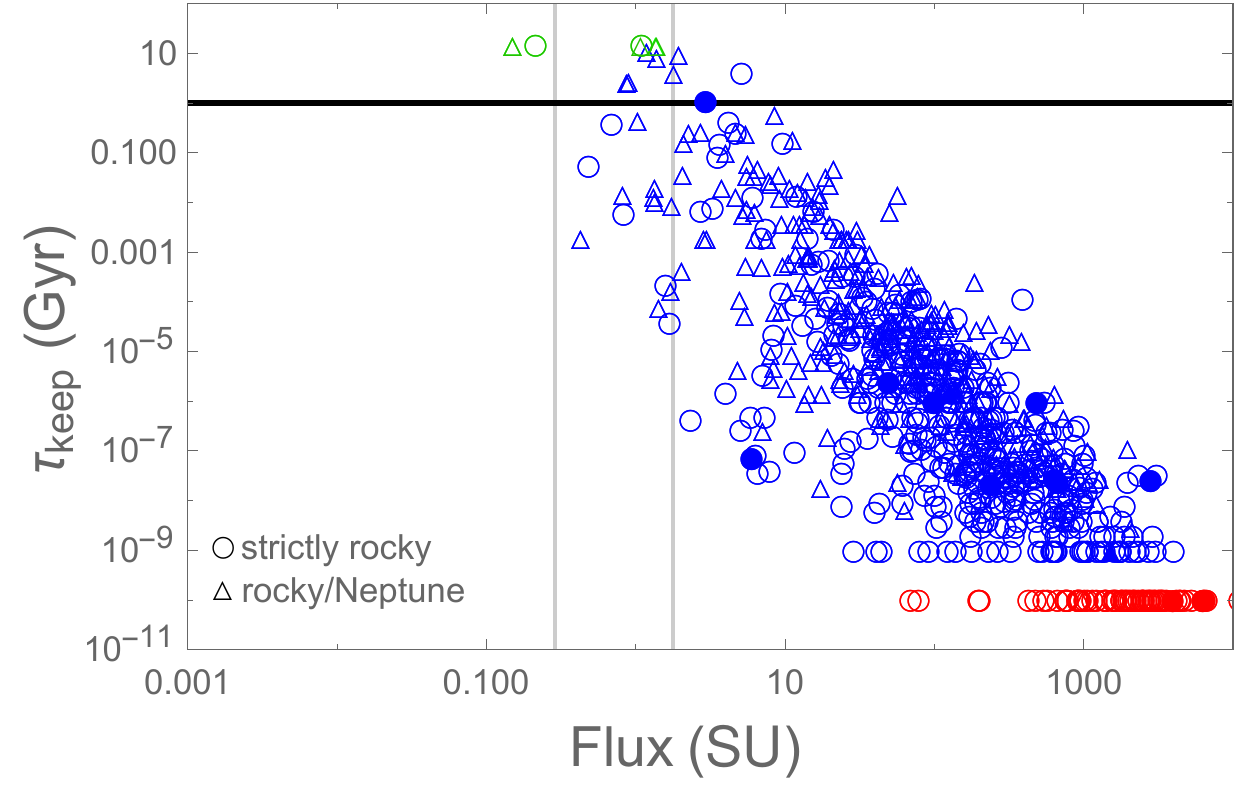}
\caption{Similar to Figure~\ref{rockycpl} except using the CTL model. The 3 planets in the rectangle in Figure~\ref{rockycpl} are now green, meaning they can hold a moon for over 15\,Gyr. Also, due to the overall longer timescales in the CTL model, there are now 7 planets in the rectangle of interest.}
\label{rockyctl}
\epsscale{1.0}
\end{figure}

In Figure~\ref{rockycpl}, we can note that there are 3 exoplanets in the rectangle we are interested in. These objects are described in more detail in the next section. Figure~\ref{rockyctl} shows the CTL model applied to the same rocky group of exoplanets. As mentioned earlier, the timescales are longer in this model (for every planet receiving less than 1.78\,SU). Indeed we now have 7 exoplanets in the rectangle and note that the 3 planets from the CPL plot have been promoted from blue to green. Past 1.78\,SU many timescales are actually lowered when moving from CPL to CTL. This is probably because these planets are very close in, so that the initial rotation rate is relatively fast leading to a large torque in the CTL model. Thus although in the CTL model the torque diminishes with time, the large initial torque is enough to compensate so that the overall timescale is larger. However, if we focus on the habitable zone rectangle, we have longer timescales for the CTL model for the planets of most interest.

\begin{figure}
\epsscale{1.2}
\plotone{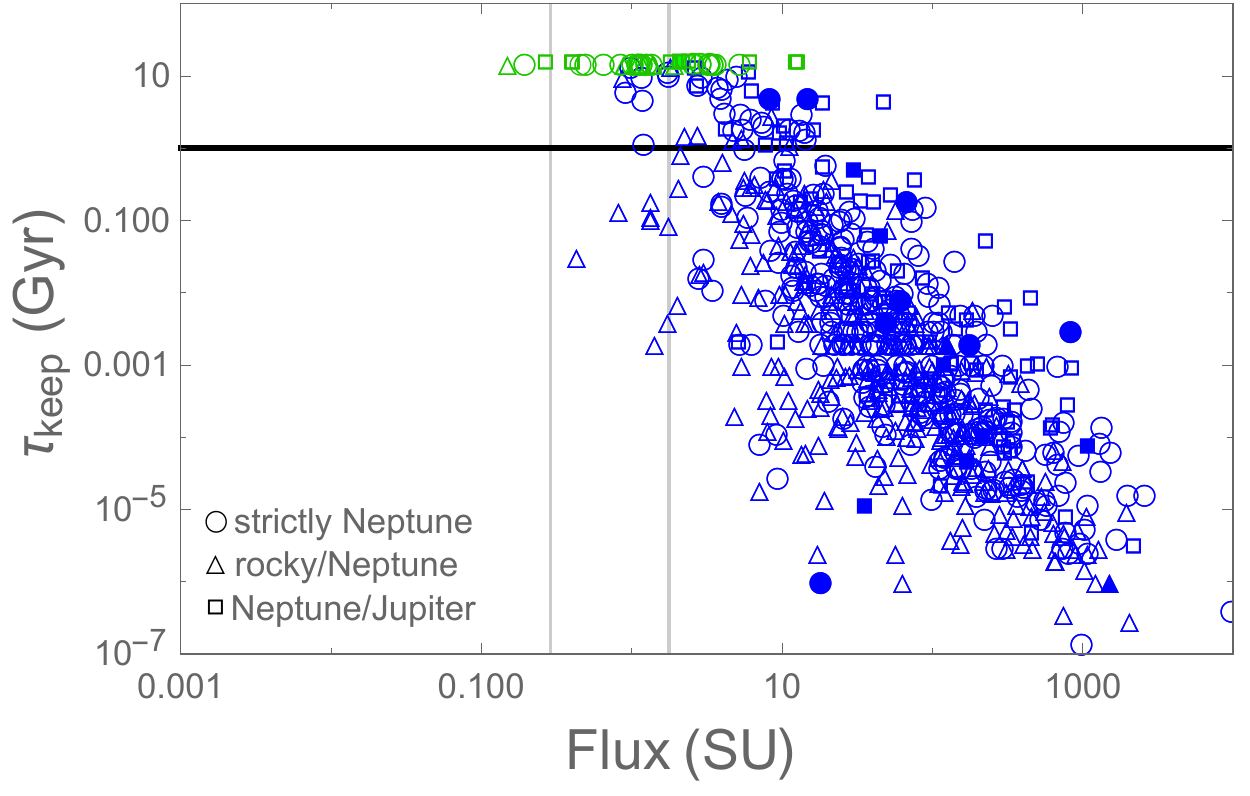}
\caption{Moon keep times for Neptune-like exoplanets using the CPL model. Circles are Neptune-like planets while triangles are treated as both Neptunes and rocky, and squares are planets taken to be both Neptune-like and Jupiter-like. Due to the less dissipative nature of Neptune-like planets versus rocky planets, there is a trend upward in moon keep time. There are a total of 87 planets that may hold a moon for over 1\,Gyr, 27 of which are in the habitable zone. (Note that the y-axis has been shifted upward, and there are some exoplanets not shown that lose their moon to their parent star).}
\label{neptunecpl}
\epsscale{1.0}
\end{figure}

\begin{figure}
\epsscale{1.2}
\plotone{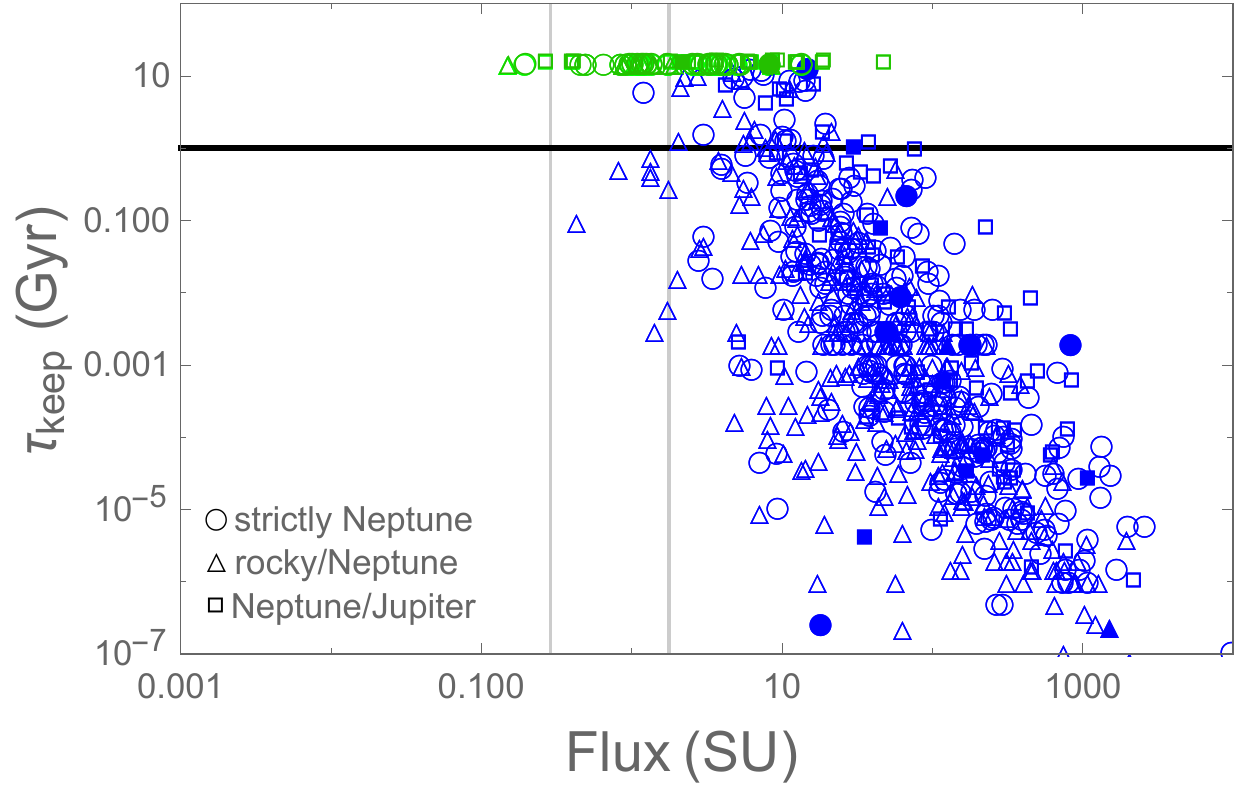}
\caption{Similar to Figure~\ref{neptunecpl}, this shows moon keep times for Neptune-like planets using the CTL model. Some of the exoplanets that are blue in Figure~\ref{neptunecpl} are green here, but the total number of planets in the rectangle is still 27 (although the total above the horizontal line has jumped to 109). }
\label{neptunectl}
\epsscale{1.0}
\end{figure}

Figures \ref{neptunecpl} and \ref{neptunectl} show the CPL and CTL model, respectively, for Neptune-like exoplanets. Because these types of planets are significantly less dissipative, we have much longer timescales overall. In fact, there are 27 exoplanets in the rectangle of interest, with most of them holding a moon for longer than 15\,Gyr. The blue triangles in Figure~\ref{rockyctl} that are exoplanets treated as both rocky and Neptune-like are all green triangles in these plots, showing that the same planet can have very different timescales based on composition. Also, many rocky planets that are in the habitable zone but have a moon survival time of less than 1\,Gyr (in between the two vertical lines but under the horizontal line in Figures \ref{rockycpl} and \ref{rockyctl}) are promoted to the rectangle of interest when treated as Neptune-like. Note the shift in the vertical axis for the Neptune-like planet: the minimum time is raised from $10^{-2}$\,yr to $10^{-7}$\,yr eliminating the red symbols (moons lost to star) that do exist. 

\begin{figure}
\epsscale{1.2}
\plotone{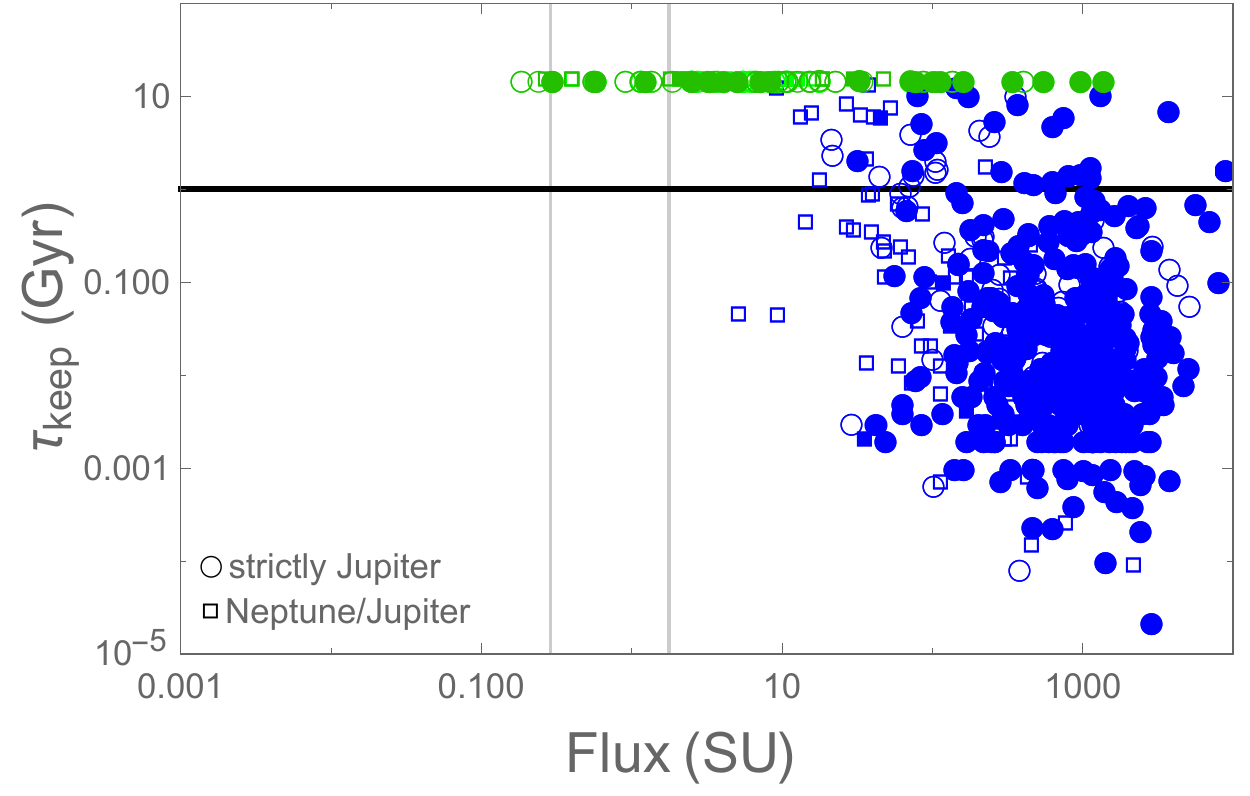}
\caption{Moon keep times for Jupiter-like exoplanets using the CPL model. Here, circles represent only Jupiters while squares are planets treated as both Jupiter-like and Neptune like. Jupiters are the least dissipative of the three types, and so have the longest timescales. There are a total of 133 planets that may hold a moon for over 1\,Gyr, but only 12 are in the habitable zone. Again, the y-axis has been shifted upward, and there are some exoplanets not shown that lose their moon to their parent star.}
\label{jupitercpl}
\epsscale{1.0}
\end{figure}

\begin{figure}
\epsscale{1.2}
\plotone{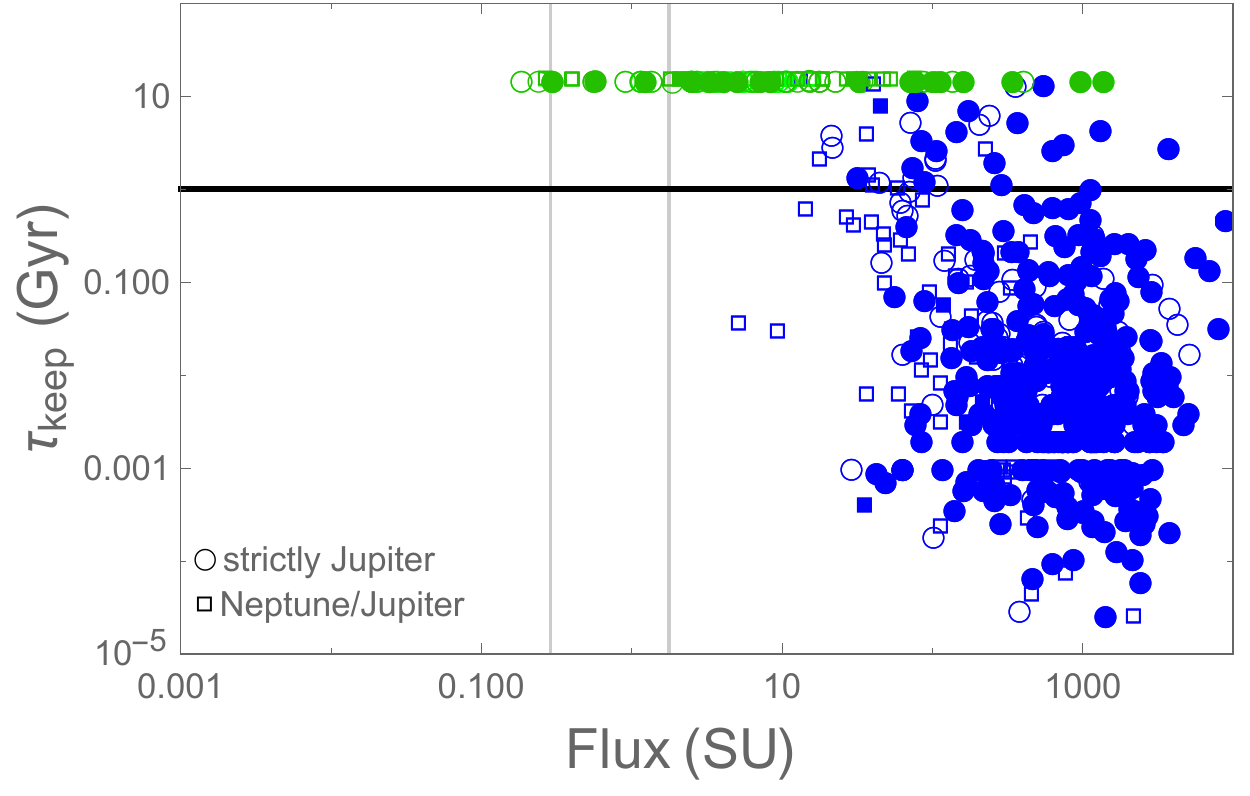}
\caption{Similar to Figure~\ref{jupitercpl}: moon keep times for Jupiter-like exoplanets using the CTL model. Overall, there is not much difference between the two models for these Jupiter-like planets, possibly because of the already very low dissipation in these systems that lead to long timescales. }
\label{jupiterctl}
\epsscale{1.0}
\end{figure}

The Jupiter-like planets are displayed in Figures \ref{jupitercpl} and \ref{jupiterctl}. Most of these planets are close in to their parent star, so we see a trend downward in timescales from CPL to CTL. However, because Jupiter-like planets are even less dissipative than Neptune-like planets, the overall timescales are longer. There are 12 planets in the rectangle of interest, all of which have a moon survival time of over 15\,Gyr. Another feature of these plots is the significant increase in filled symbols, which are planets that have a mass determination. This is expected since these are large planets whose mass is easier to measure using the radial velocity follow up measurements.

In the next section, we look into more detail at the exoplanets falling into our category of interest.

\section{Specific Systems of Interest}
\label{sec:results}

Our results show that there are at least 36 exoplanets in the habitable zone of their respective systems that can potentially hold a moon for longer than 1\,Gyr. These are summarized in Table~\ref{table2}, along with the moon survival time for each model and composition. Also included is the distance the system is from Earth, and the apparent G-band magnitude of the star.

\begin{deluxetable*}{lccccccc}
\tablecolumns{8} 
\tablewidth{200pt}
\tablecaption{Habitable Zone Planets with $\tau_{\rm{keep}}$ > 1 Gyr}
\tablehead{
Planet& Flux (SU)& Type & $\tau_{\rm{keep}}$ CPL (Gyr) & $\tau_{\rm{keep}}$ CTL (Gyr) & Distance (pc) & g-band (mag)& TESS Sector
}
\startdata
Kepler-452b & 1.104 & Rocky & 8.57 & >15 & 560 & 13.40 & S14,S15\\
Kepler-1638b & 1.382 & Rocky or Neptune & 1.60, >15 & >15, >15 & 879 & 14.75& S14,S15 \\
Kepler-22b & 1.091 & Rocky or Neptune & 5.34, >15 & >15, >15 & 190&11.64 & S15 \\
Kepler-1544b & 0.890 & Rocky or Neptune & 0.23, 15 & 2.79, >15 & 349 & 14.04 & S14,S15 \\
Kepler-1606b & 1.402 & Rocky or Neptune & 0.68, >15 & 8.55, >15 & 880 & 15.11 & S14,S26 \\
Kepler-1653b & 1.038 & Rocky or Neptune & 0.05, >15 & 0.47, >15 & 755 & NA & S14,S15 \\
Kepler-1090b & 1.200 & Rocky or Neptune & 0.91, >15 & 11.16, >15 & 879 & 14.99& S14,S15 \\
Kepler-443b & 0.880 & Rocky or Neptune & 0.29, 9.95 & 2.66, >15 & 821 & 15.90 & S15,S26 \\
Kepler-1552b & 1.100 & Neptune & >15 & >15 & 786 & 14.93 & S14 \\
Kepler-1632b & 1.270 & Neptune & >15 & >15 & 732 &13.07 &S14,S15 \\
Kepler-1540b & 0.920 & Neptune & 6.1 & >15 & 247 &14.08 & S14,S26 \\
Kepler-896b & 1.740 & Neptune & 11.5 & >15 & 896 &15.43& S14,S26 \\
Kepler-991b & 1.201 & Neptune & 1.15 & 6.1 & 394 &15.15& S14,S15 \\
Kepler-1362b & 1.181 & Neptune & 9.75 & >15 & 747 &15.82& S14 \\
Kepler-1539b & 1.770 & Neptune & 9.98 & >15 & 765 &15.31& S14,S26 \\
Kepler-1545b & 1.371 & Neptune & >15 & >15 & 745 &15.03& S14,S15 \\
Kepler-1058b & 1.202 & Neptune & 4.78 & >15 & 525 &15.5& S14,S15 \\
Kepler-1554b & 1.170 & Neptune & >15 & >15 & 973 &15.71& S14 \\
Kepler-1341b & 0.985 & Neptune & 13.43 & >15 & 483 &15.02& S14,S26 \\
Kepler-1318b & 0.495 & Neptune & >15 & >15 & 481 &15.24 &S14 \\
Kepler-1600b & 0.459 & Neptune & >15 & >15 & 1060 &15.81& S14 \\
Kepler-1593b & 1.020 & Neptune & >15 & >15 & 650 &15.88& S14,S15 \\
Kepler-1634b & 0.651 & Neptune & >15 & >15 & 628 &14.49& S14,S15 \\
Kepler-1636b & 0.850 & Neptune & >15 & >15 & 2062 &15.79& S14,S15 \\
Kepler-1097b & 1.141 & Neptune & >15 & >15 & 735 &15.21& S14,S26 \\
Kepler-1635b & 0.401 & Neptune or Jupiter & >15, >15 & >15, >15 & 1100 &15.69& S14,S26 \\
Kepler-459b & 0.400 & Neptune or Jupiter & >15, >15 & >15, >15 & 1525&15.30 & S14,S15,S26 \\
Kepler-1625b & 1.160 & Jupiter & >15 & >15 & 2460 &15.76& S14,S15 \\
Kepler-453b & 0.921 & Jupiter & >15 & >15 & 449 &13.52& S14,S26 \\
Kepler-1628b & 0.560 & Jupiter & >15 & >15 & 352 &16.95& S14,S26 \\
Kepler-456b & 0.581 & Jupiter & >15 & >15 & 758 &12.69& S14 \\
Kepler-1519b & 1.361 & Jupiter & >15 & >15 & 1503 &14.80& S14,S15 \\
Kepler-34b & 1.250 & Jupiter & >15 & >15 & 1897 &14.83& S14,S15 \\
Kepler-1654b & 0.298 & Jupiter & >15 & >15 & 578 &13.41& S14,S26 \\
Kepler-86b & 1.147 & Jupiter & >15 & >15 & 347 &12.58& S14,S15,S26 \\
Kepler-1647b & 0.575 & Jupiter & >15 & >15 & 1256 &13.58& S14,S15 
\enddata
\tablenotetext{}{}
\label{table2}
\end{deluxetable*}

\textit{Kepler-22b}. The closest system with a possible Earth-like planet is Kepler-22, at 587 light years. Kepler-22b has been an object of interest since its discovery in 2011, as it was the first transiting planet found to be in the habitable zone \citep{kepler22b}. If the planet is of rocky composition, it can hold a moon for just over 5\,Gyr using the CPL model and for over 15\,Gyr using the CTL model. If it is Neptune-like, it can potentially hold a moon for over 15\,Gyr using either tidal lag model. The system is about 4\,Gyr old \citep{Safonova}, so regardless of the composition or model, it is likely that if Kepler-22b formed an exomoon, it would retain it until the present day. In fact, the Hunt for Exomoons with Kepler (HEK) project has studied Kepler-22b and constrained a satellite mass to less than $0.5 M_{\oplus}$ with 95\% confidence \citep{kippingK22}.

\textit{Kepler-1638b}. Another notable potentially rocky exoplanet is Kepler-1638b, discovered in 2016 \citep{kepler1638b}. With a stellar flux of 1.038 SU, it is the most similar to Earth in terms of radiation received. However, if it is a rocky planet it can potentially hold a moon for only 51\,Myr or 468\,Myr, in the CPL and CTL model respectively. On the other hand, if it is actually a sub-Neptune, the moon survival time jumps to over 15\,Gyr for either model. A relatively old system, it is estimated to be around 7.7\,Gyr, so if a moon is found orbiting Kepler-1638b it is a strong indication that the planet is not actually rocky.

\textit{Kepler-991b}. Kepler-991b, also discovered in 2016 \citep{kepler1638b}, is a Neptune-like planet with a large variation in moon survivability times across the two tidal lag models. In the CPL model, this time is 1.15\,Gyr whereas in the CTL model it is 6.101\,Gyr. The age of the system is estimated to be around 3.47\,Gyr. Thus, should a moon be found around Kepler-991b it would provide a better constraint on the tidal dissipation for this planet.

\textit{Kepler-1635b}. A notable exoplanet we consider as both Neptune-like and Jupiter-like is Kepler-1635b. For both compositions and tidal dissipation models, this planet has the potential to hold a moon for longer than 15\,Gyr. This makes it a solid target for follow-up observations for exomoons in the future. In addition, experimenting with moon size and initial rotation rates in our analysis would separate timescales, with Jupiter-like having the longer timescale. Then a good constraint on the age of the system may reveal which type of planet it is. Either way, Kepler-1635b is certainly gaseous, so it is a good candidate for possibly finding life on an exomoon.

\begin{figure}
\epsscale{1.2}
\plotone{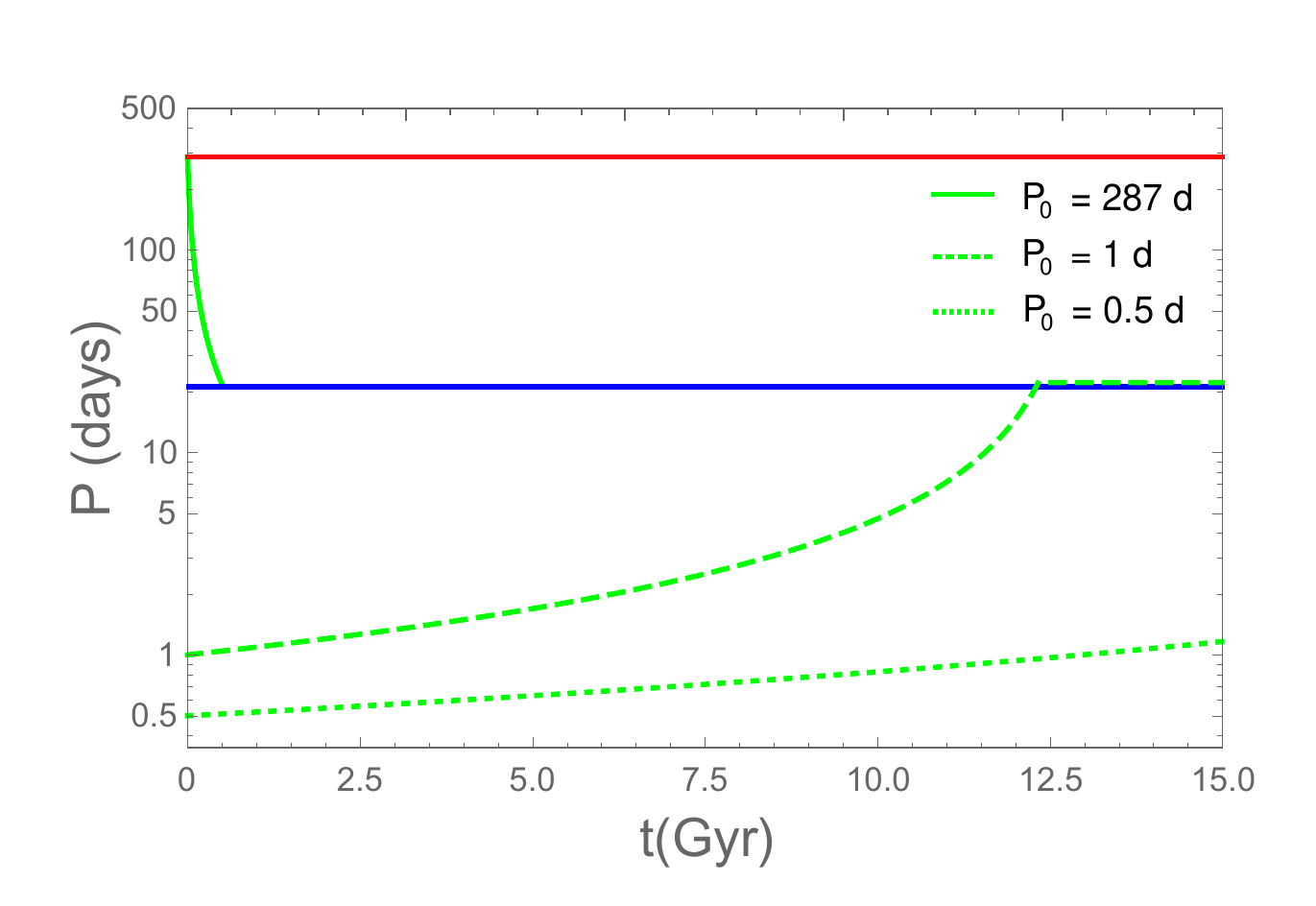}
\caption{Orbital and spin periods for the Kepler-1625 system using a Neptune-sized moon. The blue and red curves represent the the orbital period of the moon around Kepler-1625b and Kepler-1625b around its star, respectively. The green curves are the spin period of the planet. For each of the 3 initial spin rates of the planet shown, the planet itself spins up or down significantly over 15\,Gyr, but the moon's orbit stays relatively fixed.}
\label{kepler1625}
\epsscale{1.0}
\end{figure}

\textit{Kepler-1625b}. An interesting case is the Jupiter-like exoplanet Kepler-1625b, the only exoplanet (at the time of writing) to host an exomoon candidate \citep{teachey}. It is in or near the habitable zone and can potentially hold a moon for over 15\,Gyr when using a Moon sized moon. In fact, the detected exomoon is estimated to be Neptune-sized, bringing the system closer to being `binary planets' as discussed in \citet{schneider}. When running similar calculations, we find that a Neptune-sized moon also stays for over 15\,Gyr, regardless of the initial rotational period of the planet, showing that such a moon would be stable for a long time. Figure~\ref{kepler1625} shows the orbital evolution of the system using 3 different initial planet rotational periods: 287 days (initially locked to star), 1 day, and 12 hours (similar to Jupiter). In each of these cases, the moon orbit has no significant evolution, signifying the long-term retention of such a moon around this planet. Given the stability of the system, this exomoon may even host a submoon, as shown in \citet{kollmeier}, but also see \citet{rosario}. Thus, although there is much controversy surrounding the existence of this exomoon (see \citealp{kreidberg}, \citealp{heller} , and \citealp{kai} for instance), its apparent detection is a step forward into the future of exomoon discoveries.

\section{Future Implications}
\label{sec:future}

The results of our study can influence future research in several ways. First, we provide numerous targets that are probable places for exomoons to exist. We have pointed out 36 such planets, all of which are in the habitable zone of their respective systems and can hold a moon for longer than 1\,Gyr. Thus, as technology advances and observing power increases, observers can refer to our list for potential targets.

Second, because we have allowed for overlap between planet types, our work can help constrain planet composition in the event of a future detection of an exomoon. In fact, the discovery of an exomoon would eliminate at least one of our many parameters, as we have generically added a moon identical to ours in each of these systems. Reworking our analysis would then provide more accurate results.

Finally, this work adds to our understanding of tidal evolution and specifically the two current tidal lag models. Since in many of our cases the CPL and CTL models predict very different outcomes, future observations may support one of these models over the other. 

In this paper we considered few of many parameters. As shown in \citet{Piro18}, initial rotational frequency has a significant impact on moon survival time: shortening the spin period of a planet can dramatically increase this timescale. Also, as mentioned earlier, hot Jupiters can affect their host star's spin rate so that the approximation in Equation~\ref{seps} breaks down. Thus, including different planet spin periods and the star's spin can be explored in future work.

One of the biggest constraints on our models is that we consider only single-planet systems. This ensures we have simple coupled differential equations that give clean results. However, this also cuts out over 2000 exoplanets, many of which are in the habitable zone of their systems. Thus, a future project can be an extension of this work to multi-planet systems. This would take a reconfiguration of our differential equations, or we may need to turn to more advanced programs should that prove too complex.

In addition to number of planets, there are several other factors that can affect the moon retention timescale that we do not consider here. This includes the possible migration of a planet-moon system inwards leading to a smaller $a_{m,\mathrm{max}}$ and potential moon loss. The same effect may send the moon into an evection resonance orbit and collision course towards the planet \citep{Spalding}. These issues, among other complex effects, should be considered in future work.

Finally, an exciting part of exoplanet research is the observations currently being made by the Transiting Exoplanet Survey Satellite (TESS), which will view 200,000 nearby stars and catalog thousands of planets. The new discoveries that TESS makes will add to the growing list of habitable zone planets and to our list of potential exomoon hosts as well. With ground-based follow up observations, exoplanet mass will be much better constrained. This has a significant impact on our work since most of the planets we have analyzed do not have a mass estimation. Thus, in the future we will work to include many more planets in our study and incorporate newly discovered constraints, increasing the number of exoplanets potentially holding an exomoon and improving the accuracy of our results.

TESS observations of the 36 cataloged planets, noted by which sector the planet is viewed in, are included in Table~\ref{table2}.

A current project working on detecting exomoons is HEK, which uses transit timing variations (TTV) and radial velocity methods to search for exomoons \citep{hek}, and has detected the exomoon candidate around Kepler-1625b. Although there are no confirmed cases, HEK sets the precedence for using TTV's for exomoon detection, and future missions such as ESA's ARIEL (Atmospheric Remote-sensing Infrared Exoplanet Large-survey) and CHEOPS (CHaracterising ExOPlanets Satellite) will utilize transits to high precision when looking for exoplanets and possibly exomoons (\citealp{ariel,cheops}).

A closely related technique is using transit duration variation (TDV), which means how long the transit lasts varies from one orbit to another, signaling the presence of an exomoon or another planet \citep{kipping}. This technique can be utilized by TESS, ARIEL, CHEOPS, and the planned PLATO (PLAnetary Transits and Oscillations of stars) satellite mission \citep{plato}.

Additional methods proposed in finding exomoons include direct imaging with spectroastrometry \citep{agol}, which may be possible with the James Webb Space Telescope (JWST) using instruments specializing in high-contrast imaging \citep{boccaletti}. A novel idea for searching for exomoons is by following Na I and K I signatures, which are signs of a geologically active satellite \citep{oza}. Although the method is mainly for hot Jupiters in ambient plasma and does not apply directly to the habitable zone exoplanets, our study does include Kepler-14b, a hot Jupiter that can potentially hold a moon for over 15\,Gyr, and thus may be a good target to apply this approach. Finally, exomoons may also be detected using microlensing. \citet{liebig} describes the feasibility of such a process using a single planet-moon-star system, similar to the configuration chosen for our analysis. This method may be possible with PLATO, although it will specialize in transit techniques. Overall, the number of missions and methods aimed at discovering exoplanets and exomoons make the prospects of a confirmed exomoon a near-future possibility (also see the discussion in \citealp{sucerquia}.)

\section{Conclusion}
\label{sec:conclusion}
In this work, we have explored the potential for exoplanets to host exomoons. We first introduced the coupled differential equations that govern the orbital and rotational evolution of planet-star-moon systems. This includes the fact that a moon can only be stable in a limited range of separations from its parent planet. We detailed the two tidal lag models: CPL and CTL, and applied them to an Earth-like system to show the similarity in final result and difference in timescale.

Next, we analyzed theoretical systems. We first showed an example in which a planet initially tidally locked to its star can be ``unlocked'' due to the presence of a moon by considering a system with and without a moon. When calculating moon survival time, the separations $a_s$ and $a_m$ are most influential, but we showed that the planet radius is probably the next most significant parameter. We accomplished this by making a series of contour plots that displayed the trends with different parameters of the systems.

We then applied our models to real data by taking single-planet systems from the NASA Exoplanet Archive. We defined our own planet classification scheme and split exoplanets into rocky, Neptune-like, and Jupiter-like. Our main results are Figures \ref{rockycpl} through \ref{jupiterctl}, which show $\tau_{\mathrm{keep}}$ versus incident flux on the planet, and the area of interest is a moon survival time of over 1\,Gyr and stellar flux such that the planet is in the habitable zone. We found a total of 36 unique exoplanets that follow this criteria: 1 solely rocky, 7 rocky and Neptune-like, 18 solely Neptune-like, 2 Neptune-like and Jupiter-like, and 10 solely Jupiter-like. Thus we find that some of the best locations to host life in the galaxy may be the moons of gas giants.

Notable results we discussed include Kepler-22b: a possible rocky planet that is a good target to search for an exomoon, Kepler-1638b: an exoplanet whose composition can be known should an orbiting moon be found, Kepler-991b: a Neptune-like planet that may help measure tidal dissipation rates, Kepler-1635b: a gaseous planet that is an excellent candidate to have a moon with the potential to harbor life, and Kepler-1625b: an exoplanet with a potential exomoon.

In the future, the detection of moons around exoplanets may be an important constraint on exoplanet structure because the tidal dissipation and thus the time to keep a moon can vary greatly for different types of exoplanets. Such constraints can be assisted with more detailed models of the tidal dissipation since in principle the tidal reaction can be different for forcing from the moon and parent star. This should be explored in future theoretical work.

\acknowledgments
We thank David Kipping and Shreyas Vissapragada for detailed feedback. We also thank Jaime Alvarado-Montes, Jos\'{e} Caballero, Ren\'{e} Heller, Apurva Oza, Sean Raymond, Jean Schneider, and Johanna Teske for helpful comments. AT acknowledges support from the USC-Carnegie fellowship.

\bibliographystyle{yahapj}

\begin{thebibliography}{}
\providecommand\natexlab[1]{#1}
\providecommand\JournalTitle[1]{#1}

\bibitem[{Adams \& Bloch(2016)}]{adams2016}
Adams, F.~C., \& Bloch, A.~M. 2016,
  \href{http://dx.doi.org/10.1093/mnras/stw1883}{\JournalTitle{\mnras}, 462,
  2527}

\bibitem[{Agol {et~al.}(2015)Agol, Jansen, Lacy, Robinson, \& Meadows}]{agol}
Agol, E., Jansen, T., Lacy, B., Robinson, T.~D., \& Meadows, V. 2015,
  \href{http://dx.doi.org/10.1088/0004-637X/812/1/5}{\JournalTitle{\apj}, 812,
  5}
  
\bibitem[{Alvarado-Montes {et~al.}(2017)Alvarado-Montes, Zuluaga, \& Sucerquia}]{alvarado}
Alvarado-Montes, J.~A., Zuluaga, J.~I., \& Sucerquia, M. 2017,
\href{http://dx.doi.org/10.1093/mnras/stx1745}{\JournalTitle{\mnras}, 471, 3019
}
  

\bibitem[{Andrault {et~al.}(2016)Andrault, Monteux, Le~Bars, \&
  Samuel}]{andrault}
Andrault, D., Monteux, J., Le~Bars, M., \& Samuel, H. 2016,
  \href{http://dx.doi.org/10.1016/j.epsl.2016.03.020}{\JournalTitle{E\&PSL},
  443, 195}

\bibitem[{Barnes \& O’brien(2002)}]{barnes2002}
Barnes, J.~W., \& O’brien, D. 2002,
  \href{http://dx.doi.org/10.1086/341477}{\JournalTitle{\apj}, 575, 1087}

\bibitem[{Boccaletti {et~al.}(2015)Boccaletti, Lagage, Baudoz, Beichman,
  Bouchet, Cavarroc, Dubreuil, Glasse, Glauser, Hines, {et~al.}}]{boccaletti}
Boccaletti, A., Lagage, P.-O., Baudoz, P., {et~al.} 2015,
  \href{http://dx.doi.org/10.1086/682256}{\JournalTitle{\pasp}, 127, 633}

\bibitem[{Borucki {et~al.}(2012)Borucki, Koch, Batalha, Bryson, Rowe, Fressin,
  Torres, Caldwell, Christensen-Dalsgaard, Cochran, {et~al.}}]{kepler22b}
Borucki, W.~J., Koch, D.~G., Batalha, N., {et~al.} 2012,
  \href{http://dx.doi.org/10.1088/0004-637X/745/2/120}{\JournalTitle{\apj},
  745, 120}
  
\bibitem[{Cabrera \& Schneider(2007)}]{schneider}
Cabrera, J., \& Schneider, J. 2007,
  \href{http://dx.doi.org/10.1051/0004-6361:20066111}{\JournalTitle{\aap}, 464,
  1133}

\bibitem[{Cessa {et~al.}(2017)Cessa, Beck, Benz, Broeg, Ehrenreich, Fortier,
  Peter, Magrin, Pagano, Plesseria, Steller, Szoke, Thomas, Ragazzoni, \&
  Wildi}]{cheops}
Cessa, V., Beck, T., Benz, W., {et~al.} 2017,
  \href{http://dx.doi.org/10.1117/12.2304164}{in SPIE Conf. Ser., Vol. 10563},
  468

\bibitem[{Charbonneau {et~al.}(2009)Charbonneau, Berta, Irwin, Burke, Nutzman,
  Buchhave, Lovis, Bonfils, Latham, Udry, {et~al.}}]{charbonneau2009}
Charbonneau, D., Berta, Z.~K., Irwin, J., {et~al.} 2009,
  \href{http://dx.doi.org/10.1038/nature08679}{\JournalTitle{\nat}, 462, 891}

\bibitem[{Checlair {et~al.}(2017)Checlair, Menou, \& Abbot}]{checlair}
Checlair, J., Menou, K., \& Abbot, D.~S. 2017,
  \href{http://dx.doi.org/10.3847/1538-4357/aa80e1}{\JournalTitle{\apj}, 845,
  132}

\bibitem[{Domingos {et~al.}(2006)Domingos, Winter, \& Yokoyama}]{Domingos2006}
Domingos, R., Winter, O., \& Yokoyama, T. 2006,
  \href{http://dx.doi.org/10.1111/j.1365-2966.2006.11104.x}{\JournalTitle{\mnras},
  373, 1227}

\bibitem[{Efroimsky \& Lainey(2007)}]{Efroimsky}
Efroimsky, M., \& Lainey, V. 2007,
  \href{http://dx.doi.org/10.1029/2007JE002908}{\JournalTitle{JGRE}, 112}

\bibitem[{Efroimsky \& Makarov(2013)}]{efroimsky13}
Efroimsky, M., \& Makarov, V.~V. 2013,
  \href{http://dx.doi.org/10.1088/0004-637X/764/1/26}{\JournalTitle{\apj}, 764,
  26}

\bibitem[{Fortney {et~al.}(2007)Fortney, Marley, \& Barnes}]{fortney2007}
Fortney, J.~J., Marley, M.~S., \& Barnes, J.~W. 2007,
  \href{http://dx.doi.org/10.1086/512120}{\JournalTitle{\apj}, 659, 1661}

\bibitem[{Frank {et~al.}(2002)Frank, King, Raine, {et~al.}}]{Frank2002}
Frank, J., King, A., Raine, D., {et~al.} 2002, Accretion Power in Astrophysics
  (Cambridge University Press), 398

\bibitem[{Fulton {et~al.}(2017)Fulton, Petigura, Howard, Isaacson, Marcy,
  Cargile, Hebb, Weiss, Johnson, Morton, {et~al.}}]{fulton2017}
Fulton, B.~J., Petigura, E.~A., Howard, A.~W., {et~al.} 2017,
  \href{http://dx.doi.org/10.3847/1538-3881/aa80eb}{\JournalTitle{\aj}, 154,
  109}

\bibitem[{Gavrilov \& Zharkov(1977)}]{Gavrilov}
Gavrilov, S., \& Zharkov, V. 1977,
  \href{http://dx.doi.org/10.1016/0019-1035(77)90015-X}{\JournalTitle{\icarus},
  32, 443}

\bibitem[{Goldreich \& Soter(1966)}]{goldreich}
Goldreich, P., \& Soter, S. 1966,
  \href{http://dx.doi.org/10.1016/0019-1035(66)90051-0}{\JournalTitle{\icarus},
  5, 375}

\bibitem[{Heller {et~al.}(2019) Heller, Rodenbeck, \& Bruno}]{heller} Heller, R. Rodenbeck, K. \& Bruno, G. 2019,
\href{http://dx.doi.org/10.1051/0004-6361/201834913}{\JournalTitle{\aap}, 624, A95
}

\bibitem[{Heng {et~al.}(2014)Heng, Alibert, \& Benz}]{plato}
Heng, K., Alibert, Y., \& Benz, W. 2014,
  \href{http://dx.doi.org/10.1007/s10686-014-9383-4}{\JournalTitle{ExA}, 38,
  249}

\bibitem[{Jackson {et~al.}(2008)Jackson, Barnes, \& Greenberg}]{jackson}
Jackson, B., Barnes, R., \& Greenberg, R. 2008,
  \href{http://dx.doi.org/10.1111/j.1365-2966.2008.13868.x}{\JournalTitle{\mnras},
  391, 237}

\bibitem[{Kipping(2009)}]{kipping}
Kipping, D.~M. 2009,
  \href{http://dx.doi.org/10.1111/j.1365-2966.2008.13999.x}{\JournalTitle{\mnras},
  392, 181}

\bibitem[{Kipping {et~al.}(2012)Kipping, Bakos, Buchhave, Nesvorn{\`y}, \&
  Schmitt}]{hek}
Kipping, D.~M., Bakos, G.~{\'A}., Buchhave, L., Nesvorn{\`y}, D., \& Schmitt,
  A. 2012,
  \href{http://dx.doi.org/https://doi.org/10.1088/0004-637X/750/2/115}{\JournalTitle{\apj},
  750, 115}

\bibitem[{Kipping {et~al.}(2013)Kipping, Forgan, Hartman, Nesvorný, Bakos,
  Schmitt, \& Buchhave}]{kippingK22}
Kipping, D.~M., Forgan, D., Hartman, J., {et~al.} 2013,
  \href{http://dx.doi.org/10.1088/0004-637x/777/2/134}{\JournalTitle{\apj},
  777, 134}

\bibitem[{Kite {et~al.}(2011)Kite, Gaidos, \& Manga}]{kite}
Kite, E.~S., Gaidos, E., \& Manga, M. 2011,
  \href{http://dx.doi.org/10.1088/0004-637X/743/1/41/meta}{\JournalTitle{\apj},
  743, 41}

\bibitem[{Kollmeier \& Raymond(2018)}]{kollmeier}
Kollmeier, J.~A., \& Raymond, S.~N. 2018,
  \href{http://dx.doi.org/10.1093/mnrasl/sly219}{\JournalTitle{\mnras}, 483,
  L80}

\bibitem[{Kopparapu {et~al.}(2013)Kopparapu, Ramirez, Kasting, Eymet, Robinson,
  Mahadevan, Terrien, Domagal-Goldman, Meadows, \& Deshpande}]{kopparapu2013}
Kopparapu, R.~K., Ramirez, R., Kasting, J.~F., {et~al.} 2013,
  \href{http://dx.doi.org/10.1088/0004-637X/765/2/131}{\JournalTitle{\apj},
  765, 131}

\bibitem[{Kramm {et~al.}(2011)Kramm, Nettelmann, Redmer, \& Stevenson}]{Kramm}
Kramm, U., Nettelmann, N., Redmer, R., \& Stevenson, D.~J. 2011,
  \href{http://dx.doi.org/10.1051/0004-6361/201015803}{\JournalTitle{\aap},
  528, A18}

\bibitem[{Kreidberg {et~al.}(2019)Kreidberg, Luger, \& Bedell}]{kreidberg}
Kreidberg, L., Luger, R., \& Bedell, M. 2019,
  \href{http://dx.doi.org/10.3847/2041-8213/ab20c8}{\JournalTitle{\apjl}, 877,
  L15}

\bibitem[{Lambeck \& Runcorn(1977)}]{lambeck}
Lambeck, K., \& Runcorn, S.~K. 1977,
  \href{http://dx.doi.org/10.1098/rsta.1977.0159}{\JournalTitle{RSPTA}, 287,
  545}

\bibitem[{Laskar {et~al.}(1993)Laskar, Joutel, \& Robutel}]{laskar1993}
Laskar, J., Joutel, F., \& Robutel, P. 1993,
  \href{http://dx.doi.org/10.1038/361615a0}{\JournalTitle{\nat}, 361, 615}

\bibitem[{Liebig \& Wambsganss(2010)}]{liebig}
Liebig, C., \& Wambsganss, J. 2010,
  \href{http://dx.doi.org/10.1051/0004-6361/200913844}{\JournalTitle{\aap},
  520, A68}

\bibitem[{Mart\'{i}nez-Rodr\'{i}guez {et~al.}(2019)Mart\'{i}nez-Rodr\'{i}guez, Caballero,
  Cifuentes, Piro, \& Barnes}]{Martinez}
Mart\'{i}nez-Rodr\'{i}guez, H., Caballero, J.~A., Cifuentes, C., Piro, A.~L., \&
  Barnes, R. 2019,
  \href{http://dx.doi.org/10.3847/1538-4357/ab5640}{\JournalTitle{\apj}, 887,
  261}

\bibitem[{Morton {et~al.}(2016)Morton, Bryson, Coughlin, Rowe, Ravichandran,
  Petigura, Haas, \& Batalha}]{kepler1638b}
Morton, T.~D., Bryson, S.~T., Coughlin, J.~L., {et~al.} 2016,
  \href{http://dx.doi.org/10.3847/0004-637X/822/2/86}{\JournalTitle{\apj}, 822,
  86}

\bibitem[{Murray \& Dermott(1999)}]{murray}
Murray, C.~D., \& Dermott, S.~F. 1999, Solar system dynamics (Cambridge
  university press)

\bibitem[{Neron~de Surgy \& Laskar(1997)}]{neron}
Neron~de Surgy, O., \& Laskar, J. 1997, \JournalTitle{\aap}, 318, 975

\bibitem[{Odrzywolek \& Rafelski(2016)}]{odrzywolek2016}
Odrzywolek, A., \& Rafelski, J. 2016,
  \href{http://dx.doi.org/10.5506/APhysPolB.49.1917}{\JournalTitle{AcPPB}, 49}

\bibitem[{Ogilvie(2014)}]{Ogilvie}
Ogilvie, G.~I. 2014,
  \href{http://dx.doi.org/10.1146/annurev-astro-081913-035941}{\JournalTitle{\araa},
  52, 171}
  
\bibitem[{Oza {et~al.}(2019)Oza, Johnson, Lellouch, Schmidt, Schneider, Huang, Gamborino, Gebek, Wyttenbach, Demory}]{oza}
Oza, A.~V., Johnson, R.~E., Lellouch, E., {et~al.} 2019,
\href{http://dx.doi.org/10.3847/1538-4357/ab40cc}{\JournalTitle{\apj},
885, 168}


\bibitem[{Pascale {et~al.}(2018)Pascale, Bezawada, Barstow, Beaulieu, Bowles,
  du~Foresto, Coustenis, Decin, Drossart, Eccleston, Encrenaz, Forget, Griffin,
  Güdel, Hartogh, Heske, Lagage, Leconte, Malaguti, Micela, Middleton, Min,
  Moneti, Morales, Mugnai, Ollivier, Pace, Papageorgiou, Pilbratt, Puig, Rataj,
  Ray, Ribas, Rocchetto, Sarkar, Selsis, Taylor, Tennyson, Tinetti, Turrini,
  Vandenbussche, Venot, Waldmann, Wolkenberg, Wright, Osorio, \&
  Zingales}]{ariel}
Pascale, E., Bezawada, N., Barstow, J., {et~al.} 2018,
  \href{http://dx.doi.org/10.1117/12.2311838}{in SPIE, Vol. 10698}, 169

\bibitem[{Piro(2018)}]{Piro18}
Piro, A. 2018,
  \href{http://dx.doi.org/10.3847/1538-3881/aaca38}{\JournalTitle{\aj}, 156}
  
\bibitem[{{Rodenbeck} {et~al.}(2018){Rodenbeck},{Heller},\&{Hippke}}]{kai}
{Rodenbeck}, K., {Heller}, R., \& {Hippke}, M. 2018,
\href{http://dx.doi.org/10.1051/0004-6361/201833085}{\JournalTitle{\aap}, 617, A49
}


\bibitem[{{Rosario-Franco} {et~al.}(2020){Rosario-Franco}, {Quarles},
  {Musielak}, \& {Cuntz}}]{rosario}
{Rosario-Franco}, M., {Quarles}, B., {Musielak}, Z.~E., \& {Cuntz}, M. 2020,
  \href{http://dx.doi.org/10.3847/1538-3881/ab89a7}{\JournalTitle{\aj}, 159,
  260}

\bibitem[{Safonova {et~al.}(2016)Safonova, Murthy, \& Shchekinov}]{Safonova}
Safonova, M., Murthy, J., \& Shchekinov, Y.~A. 2016,
  \href{http://dx.doi.org/10.1017/S1473550415000208}{\JournalTitle{IJAsB}, 15,
  93}

\bibitem[{Sasaki \& Barnes(2014)}]{sasaki14}
Sasaki, T., \& Barnes, J.~W. 2014,
  \href{http://dx.doi.org/10.1017/S1473550414000184}{\JournalTitle{IJAsB}, 13,
  324}

\bibitem[{Sasaki {et~al.}(2012)Sasaki, Barnes, \& O'Brien}]{sasaki12}
Sasaki, T., Barnes, J.~W., \& O'Brien, D.~P. 2012,
  \href{http://dx.doi.org/10.1088/0004-637X/754/1/51}{\JournalTitle{\apj}, 754,
  51}

\bibitem[{Schopf {et~al.}(2018)Schopf, Kitajima, Spicuzza, Kudryavtsev, \&
  Valley}]{Schopf2017}
Schopf, J.~W., Kitajima, K., Spicuzza, M.~J., Kudryavtsev, A.~B., \& Valley,
  J.~W. 2018,
  \href{http://dx.doi.org/10.1073/pnas.1718063115}{\JournalTitle{PNAS}, 115,
  53}
  
\bibitem[{Spalding {et~al.}(2016)Spalding, Batygin, \& Adams}]{Spalding} Spalding, C. , Batygin, K. , \& Adams, F.~C. 2016, \href{http://dx.doi.org/10.3847/0004-637X/817/1/18}{\JournalTitle{\apj}, 817, 18}  
  

\bibitem[{Stern \& Levison(2002)}]{stern2002}
Stern, S.~A., \& Levison, H.~F. 2002,
  \href{http://dx.doi.org/10.1017/S1539299600013289}{\JournalTitle{HiA}, 12,
  205}

\bibitem[{Storch \& Lai(2014)}]{storch}
Storch, N.~I., \& Lai, D. 2014,
  \href{http://dx.doi.org/10.1093/mnras/stt2292}{\JournalTitle{\mnras}, 438,
  1526}

\bibitem[{{Sucerquia} {et~al.}(2020){Sucerquia}, {Ram{\'\i}rez},
  {Alvarado-Montes}, \& {Zuluaga}}]{sucerquia}
{Sucerquia}, M., {Ram{\'\i}rez}, V., {Alvarado-Montes}, J.~A., \& {Zuluaga},
  J.~I. 2020,
  \href{http://dx.doi.org/10.1093/mnras/stz3548}{\JournalTitle{\mnras}, 492,
  3499}

\bibitem[{Teachey \& Kipping(2018)}]{teachey}
Teachey, A., \& Kipping, D.~M. 2018,
  \href{http://dx.doi.org/10.1126/sciadv.aav1784}{\JournalTitle{SciA}, 4}

\bibitem[{Touma \& Wisdom(1994)}]{touma}
Touma, J., \& Wisdom, J. 1994,
  \href{http://dx.doi.org/10.1086/300312}{\JournalTitle{\aj}, 108, 1943}

\bibitem[{Valencia {et~al.}(2007)Valencia, Sasselov, \&
  O'Connell}]{valencia2007}
Valencia, D., Sasselov, D.~D., \& O'Connell, R.~J. 2007,
  \href{http://dx.doi.org/10.1086/509800}{\JournalTitle{\apj}, 656, 545}

\bibitem[{Vick {et~al.}(2019)Vick, Lai, \& Anderson}]{vick}
Vick, M., Lai, D., \& Anderson, K.~R. 2019,
  \href{http://dx.doi.org/10.1093/mnras/stz354}{\JournalTitle{\mnras}}

\bibitem[{Ward \& Reid(1973)}]{ward}
Ward, W.~R., \& Reid, M.~J. 1973,
  \href{http://dx.doi.org/10.1093/mnras/164.1.21}{\JournalTitle{\mnras}, 164,
  21}

\bibitem[{Yoder(1995)}]{Yoder1995}
Yoder, C.~F. 1995 (Wiley Online Library)

\end{thebibliography}

\end{document}